\documentclass[11pt]{article}

\usepackage[utf8]{inputenc}
\usepackage{charter}
\usepackage{fullpage}
\usepackage{hyperref}
\usepackage{graphicx}
\usepackage{lipsum}
\usepackage[sort&compress,numbers]{natbib}
\usepackage[pdftex]{color}
\usepackage{hyperref}
\hypersetup{hidelinks}
\usepackage[total={6.5in,9in},top=1in,headsep=0.1in,headheight=1in]{geometry}
\usepackage{amsmath,amssymb}

\newcommand\snowmass{
\begin{center}
  \rule[-0.2in]{\hsize}{0.01in}\\
  \rule{\hsize}{0.01in}\\
  \vskip 0.1in
  Submitted to the Proceedings of the US Community Study\\ 
  on the Future of Particle Physics (Snowmass 2021)\\
  \rule{\hsize}{0.01in}\\
  \rule[+0.2in]{\hsize}{0.01in}\\[-2em]
\end{center}
}

\definecolor{red}{rgb}{0.8,0.0,0.0}
\definecolor{green}{rgb}{0.0,0.6,0.0}
\definecolor{darkblue}{rgb}{0.0,0.1,0.7}
\definecolor{brown}{rgb}{0.6,0.1,0.0}
\definecolor{grey}{rgb}{0.6,0.6,0.6}
\definecolor{darkgreen}{rgb}{0.0, 0.545098, 0.0}
\definecolor{applegreen}{rgb}{0.55, 0.71, 0.0}
\definecolor{purple}{rgb}{0.5,0.0,0.5}
\definecolor{babypink} {rgb}{0.64, 0.44, 0.44}
\definecolor{orange}{rgb}{1.0,0.5,0.0}

\usepackage[firstpage=true]{background}
\backgroundsetup{contents={\parbox{6.5in}{\snowmass}}, scale=1,placement=top,opacity=1,color=black,position={3.25in,1.2in}}

\usepackage{fancyhdr}
\fancypagestyle{plain}{%
  \fancyhf{}%
  \fancyhead[C]{}
  \fancyfoot[C]{\thepage}
}

\fancypagestyle{empty}{%
  \fancyhf{}%
  \fancyhead[C]{{\it 
  Report of the Snowmass 2021 
  Topical Group on
Lattice Gauge Theory}}
  \fancyfoot[C]{\thepage}
}
\title{{\small Report of the Snowmass 2021 
Topical Group on}
\\
Lattice Gauge Theory}
\date{}   

\usepackage{authblk}

\author[1,2]{Zohreh~Davoudi (editor/topical-group~convener)\thanks{\tt davoudi@umd.edu}}
\author[3]{Ethan~T.~Neil (editor/topical-group~convener)\thanks{\tt ethan.neil@colorado.edu}}

\author[4]{\and Christian~W.~Bauer}
\author[5]{Tanmoy~Bhattacharya}
\author[6]{Thomas~Blum}
\author[7]{Peter~Boyle}
\author[8]{Richard~C.~Brower}
\author[9]{Simon~Catterall}
\author[10]{Norman~H.~Christ}
\author[11]{Vincenzo~Cirigliano}
\author[12]{Gilberto~Colangelo}
\author[13]{Carleton~DeTar}
\author[14,15]{William~Detmold}
\author[16]{Robert~G.~Edwards}
\author[17]{Aida~X.~El-Khadra}
\author[18]{Steven~Gottlieb}
\author[5]{Rajan~Gupta}
\author[14]{Daniel~C.~Hackett}
\author[3]{Anna~Hasenfratz}
\author[7,19]{Taku~Izubuchi~(topical-group~convener)}
\author[14]{William~I.~Jay}
\author[6]{Luchang~Jin}
\author[20]{Christopher~Kelly}
\author[21]{Andreas~S.~Kronfeld}
\author[22]{Christoph~Lehner}
\author[23,24]{Huey-Wen~Lin}
\author[20]{Meifeng~Lin}
\author[17]{Andrew~T.~Lytle}
\author[25]{Stefan~Meinel}
\author[26]{Yannick~Meurice}
\author[7]{Swagato~Mukherjee}
\author[27]{Amy~Nicholson}
\author[28,29]{Sasa~Prelovsek}
\author[30]{Martin~J.~Savage}
\author[14,15]{Phiala~E.~Shanahan}
\author[21]{Ruth~S.~Van~De~Water}
\author[21]{Michael~L.~Wagman}
\author[31]{Oliver~Witzel}

\affil[1]{Maryland Center for Fundamental Physics and Department of Physics, University of Maryland, College Park, MD 20742, USA}
\affil[2]{The NSF Institute for Robust Quantum Simulation, University of Maryland, College Park, MD 20742, USA}
\affil[3]{Department of Physics, University of Colorado, Boulder, CO 80309, USA}
\affil[4]{Physics Division, Lawrence Berkeley National Laboratory, Berkeley, CA 94720, USA}
\affil[5]{T-2, Los Alamos National Laboratory, Los Alamos, NM 87545, USA}
\affil[6]{Physics Department, University of Connecticut, Storrs, CT 06269, USA}
\affil[7]{Physics Department, Brookhaven National Laboratory, Upton, NY 11973, USA}
\affil[8]{Department of Physics and Center for Computational Science, Boston University, Boston, MA 02215, USA}
\affil[9]{Department of Physics, Syracuse University, Syracuse, NY 13244, USA} 
\affil[10]{Physics Department, Columbia University, New York, New York 10027, USA}
\affil[11]{Institute for Nuclear Theory (INT), University of Washington, Seattle WA 91195, USA}
\affil[12]{Albert Einstein Center for Fundamental Physics, Institute for Theoretical Physics, University of Bern, Bern, Switzerland}
\affil[13]{Department of Physics and Astronomy, University of Utah, Salt Lake City, UT 84112, USA}
\affil[14]{Center for Theoretical Physics, Massachusetts Institute of Technology, Boston, MA 02139, USA}
\affil[15]{The NSF AI Institute for Artificial Intelligence and Fundamental Interactions}
\affil[16]{Thomas Jefferson National Accelerator Facility, 12000 Jefferson Avenue, Newport News, VA 23606, USA}
\affil[17]{Department of Physics and Illinois Center for Advanced Studies of the Universe, University of Illinois, Urbana, IL 61801, USA}
\affil[18]{Department of Physics, Indiana University, Bloomington, Indiana 47405, USA}
\affil[19]{RIKEN-BNL Research Center, Brookhaven National Laboratory, Upton, NY 11973, USA}
\affil[20]{Computational Science Initiative, Brookhaven National Laboratory, Upton, NY 11973, USA} 
\affil[21]{Theory Division, Fermi National Accelerator Laboratory, Batavia, IL 60510, USA}
\affil[22]{Universit\"at Regensburg, Fakult\"at f\"ur Physik, 93040 Regensburg, Germany}
\affil[23]{Department of Physics and Astronomy, Michigan State University, East Lansing, MI 48824, USA}
\affil[24]{Department of Computational Mathematics, Science and Engineering, Michigan State University, East Lansing, MI 48824, USA}
\affil[25]{Department of Physics, University of Arizona, Tucson, AZ 85721, USA}
\affil[26]{Department of Physics and Astronomy, University of Iowa, Iowa City, IA 52242, USA}
\affil[27]{Department of Physics and Astronomy, University of North Carolina, Chapel Hill, NC 27516, USA}
\affil[28]{Faculty of Mathematics and Physics, University of Ljubljana, Ljubljana, Slovenia}
\affil[29]{Jozef Stefan Institute, Ljubljana, Slovenia}
\affil[30]{InQubator for Quantum Simulation (IQuS), Department of Physics,
University of Washington, Seattle, WA 98195, USA}
\affil[31]{Center for Particle Physics Siegen, Theoretische Physik 1, Universit\"at Siegen, 57068 Siegen, Germany}

\begin{document}

\maketitle

\pagebreak

\begin{abstract}
\noindent
Lattice gauge theory continues to be a powerful theoretical and computational approach to simulating strongly interacting quantum field theories, whose applications permeate almost all disciplines of modern-day research in High-Energy Physics.  Whether it is to enable precision quark- and lepton-flavor physics, to uncover signals of new physics in nucleons and nuclei, to elucidate hadron structure and spectrum, to serve as a numerical laboratory to reach beyond the Standard Model, or to invent and improve state-of-the-art computational paradigms, the lattice-gauge-theory program is in a prime position to impact the course of developments and enhance discovery potential of a vibrant experimental program in High-Energy Physics over the coming decade. This projection is based on abundant successful results that have emerged using lattice gauge theory over the years: on continued improvement in theoretical frameworks and algorithmic suits; on the forthcoming transition into the exascale era of high-performance computing; and on a skillful, dedicated, and organized community of lattice gauge theorists in the U.S. and worldwide. The prospects of this effort in pushing the frontiers of research in High-Energy Physics have recently been studied within the U.S. decadal Particle Physics Planning Exercise (\emph{Snowmass2021}), and the conclusions are summarized in this Topical Report.

\
\

\noindent
\textbf{Preprint Report No.} UMD-PP-022-08, LA-UR-22-29361, FERMILAB-CONF-22-703-T.

\vspace{0.5 cm}

\end{abstract}

\pagebreak

\tableofcontents

\pagebreak
\phantomsection
\addcontentsline{toc}{section}{Executive Summary}
\section*{Executive Summary}
Quantum field theories permeate our theoretical descriptions of Nature at the smallest distances and the highest energies. Lattice field theory is the most reliable theoretical tool to date which rigorously defines and, in a systematically improvable fashion, allows simulation of strongly interacting quantum field theories---most importantly, quantum chromodynamics (QCD), which is one of the pillars of the Standard Model.  Presently, the field of High-Energy Physics (HEP) stands at a critical point where theoretical uncertainties or lack of predictions for a range of quantities at the high-energy colliders, in searches for new-physics scenarios in rare processes, and in neutrino experiments may limit the discovery potential of these endeavors over the coming decade. Fortunately, at the same time lattice field theory has reached maturity in providing reliable and precise predictions for many of these important experiments, either directly, or indirectly via supplementing the nuclear-level computations.  With continued support and growth, lattice field theory will continue to provide these essential inputs.

In the area of quark and lepton flavor physics, precision lattice-QCD input has been and will continue to be essential for the interpretation of experimental results. Prominent examples are hadronic contributions to the muon's anomalous magnetic moment, hadronic form factors in both light- and heavy-flavor sectors that are necessary for tests of unitarity of the Cabibbo-Kobayashi-Masukawa (CKM) matrix and for validation of potential lepton-universality anomalies, as well as charge-parity (CP)-violating parameters in kaon decays.  Precision determinations of quark masses and the strong coupling constant, as well as nucleon parton distribution functions (PDFs), are required inputs for precision Higgs physics and beyond-the-Standard-Model (BSM) searches at the Large Hadron Collider (LHC).  Lattice QCD will also continue to enable the computation of nucleon/nuclear form factors and associated charges, and other nucleon matrix elements, for a variety of investigations in searches for the violation of fundamental symmetries and for new physics. Prominent examples are neutrino-oscillation experiments that are limited by uncertain neutrino-nucleus scattering cross sections, electric-dipole-moment (EDM) searches, and searches for rare processes such as neutrinoless double-$\beta$ decay, proton decay, lepton-flavor conversion in presence of nuclear media, and dark-matter--nucleus scattering. Lattice QCD further enables first-principles prediction of the exotic spectrum of QCD, and enables access to scattering amplitudes and resonance properties, which are critical for several BSM searches.

Lattice-field-theory calculations can also serve as a numerical laboratory to deepen our understanding of strongly interacting quantum field theories, both within the Standard Model and beyond.  Lattice field theory allows direct exploration of how the dynamics are altered when the fundamental parameters of the theory (such as the number of colors or the number of light fermions) are changed, leading to rigorous tests of frameworks such as the large-$N_c$ expansion.  These numerical results can also be used to constrain and inform strongly-coupled models of new physics such as composite Higgs, composite dark matter, or neutral naturalness.  Lattice explorations may also lead to dynamical surprises, such as the discovery of a light scalar resonance in the spectrum which recently led to the development of dilaton effective field theory (EFT) to describe the lattice spectrum results.  Exploration of supersymmetric theories such as $\mathcal{N} = 4$ super-Yang-Mills may give new insights into the AdS/CFT correspondence and string theory.

The projections in recent whitepapers provided by the community for the impact of lattice gauge theory in each of the areas discussed above rely on several important requirements.  With the exascale era of computing promising at least an order of magnitude increase in overall computing power over the next decade, commensurate increases in the computing resources devoted to lattice gauge theory will be necessary to ensure that the goals of the lattice gauge theory research program, as stated by the community in this planning exercise, can be achieved.  However, increased computing resources alone will not be sufficient to achieve important research goals over the next decade, as further innovations in computing algorithms and physics methods will also be necessary.  Investment in a diverse human resource with various skill sets in theory, algorithm, high-performance computing (HPC), and numerical analysis is the key in bringing new ideas and facilitating innovation, keeping up with ever-changing computing software and hardware architecture including machine learning and quantum computing, and engaging with the theoretical and experimental communities to increase impact and relevance. Supporting dedicated programs for software development and hardware integration will continue to be critical over the next decade, as is ensuring that the trained workforce will be retained in the program, e.g., by creating positions of more permanent nature, so that the continuity of the long-term projects will not be disrupted.

\section{Introduction}
The field of HEP boasts a vibrant experimental program that aims to discover new physical mechanisms at the foundations of Nature, whether through increasing energy or intensity of the probes, or through astrophysical and cosmological observations.  However, experimental progress also relies on a thriving theoretical-physics community that proposes new plausible models beyond the (accepted but incomplete) Standard Model of particle physics, and is further able to provide accurate predictions based on those models for experimental searches. Quantum field theories permeate our theoretical description of Nature at the smallest distances and the highest energies---scales that can be probed either at particle colliders of today and of the future, or indirectly through the influence of new virtual particles in low-energy experiments. Lattice field theory\footnote{Throughout this report, we use the  (more general) term ``lattice field theory'' and the (less general) term ``lattice gauge theory'' interchangeably, as well as the colloquial shorthand of simply ``lattice''.} is the most reliable theoretical tool to date to simulate strongly-interacting quantum field theories of relevance to Nature, both well-established and hypothetical.

The history of lattice field theory is a success story spanning multiple decades in developing and applying an extensive theoretical and computational suite to difficult problems in both HEP and Nuclear Physics.  This history consisted of, first of all, formally defining the field-theory path integral and various correlation functions in a finite discrete spacetime in such a way that as many symmetries as possible are kept, or systematically recovered, in the continuum infinite-volume limit, starting from the pioneering work of Wilson~\cite{wilson1974confinement} and Kogut and Susskind~\cite{kogut1975hamiltonian}. Many more theoretical and conceptual breakthroughs came along later, an example of which being the mapping that allows seemingly inaccessible few-hadron scattering amplitudes to be obtained from imaginary-time lattice field theory computations, starting from the pioneering work of  L\"uscher~\cite{Luscher:1986pf}.  The development of lattice field theory also includes devising algorithms that, over time, scaled better with the system’s parameters and took advantage not only of advances in applied mathematics and computer science but importantly of physics input, such as expression of symmetries and constraints, and renormalization group and scale separation, to make seemingly impossible computations possible. Furthermore, the success of lattice studies relied on adjusting algorithms and compilations to the hardware architecture, and remarkably in some instances, impacted the development of computing architecture itself via a co-design process~\cite{boyle2005overview,boyle2004qcdoc}. It is no surprise that the community is currently embracing new trends in computational sciences such as machine learning and quantum computing. The field of lattice field theory also enjoys a robust and organized community that generally works in harmony and shares knowledge, data, and codes to expedite science and save human- and computing-time resources.

We are standing at a critical point in HEP where the theoretical uncertainties or lack of predictions for nonperturbative quantities at the high-energy colliders, in searches for new physics in rare processes, and in neutrino experiments may limit the discovery potential of these endeavors over the coming decade. Fortunately, at the same time lattice field theory has reached the point that it is providing reliable and precise predictions for many important experiments, and is hence changing the game constantly and rapidly. Prominent examples include, but are not limited to: i) hadronic contributions to the muon's magnetic moment to confront experiment without reliance on \emph{ad hoc} models, ii) hadronic form factors in both light- and heavy-flavor sectors for tests of unitarity of the CKM matrix and for validation of potential lepton-universality anomalies, iii) CP-violating parameters in mesonic decays such as kaon decays, iv) precision calculation of quark masses and strong coupling for precision Higgs physics and BSM searches at the LHC, v) nucleon EDM, nucleon form factors, and associated charges for constraining CP violation and various BSM scenarios, for neutrino-oscillation experiments that are limited by uncertain neutrino-nucleus scattering cross sections, and for constraining dark-matter--nucleus cross sections in direct dark-matter detections, vi) nucleon PDFs and other structure functions needed for the LHC program and for augmenting the current limited phenomenological fits, vii) hadronic spectra including exotic resonances and resonance transition amplitudes, and viii) the first steps towards quantifying nuclear effects in several matrix elements of relevance to the HEP experiments. Currently, only some of these studies have reached the desired precision generally needed for the respective experiments. Nonetheless, with the availability of sufficient computing resources in the exascale era and the growth of algorithmic advances and formal understandings, the next decade we will likely witness a leap in the application of the lattice-gauge-theory methods to all these critical areas of research, as explained in this report.

Importantly, lattice methods also provide a powerful numerical laboratory to explore the nature of strongly-interacting quantum field theories, both quantum chromodynamics (QCD) and hypothetical new-physics sectors.  Strongly-coupled new physics may have properties and experimental signatures which cannot be reliably predicted using perturbative methods, and extrapolation from QCD (the only strongly-coupled gauge theory for which we have experimental data so far) may not be a reliable guide.  Lattice calculations have provided concrete predictions for composite Higgs and composite dark-matter models, narrowing the experimental parameter space where they might be found, or hinting at new dynamical mechanisms such as the emergence of a light ``pseudo-dilaton'' bound state.  Lattice calculations have tested our understanding of the larger space of field theories through constructions such as the large-$N_c$ expansion and by confirming the existence of four-dimensional systems with emergent conformal symmetry. They have also provided nonperturbative explorations of theoretical models to inform our understanding of supersymmetric field theory, string theory, and quantum gravity.  In the coming decade, continued lattice study of new strongly-coupled theories with drastically different behavior than the familiar example of QCD will require the development of new algorithms and methods, which may feed back to improve the study of QCD itself or lead to new ideas for physics beyond the Standard Model. 
\begin{figure}[t!]
\includegraphics[scale=0.945]{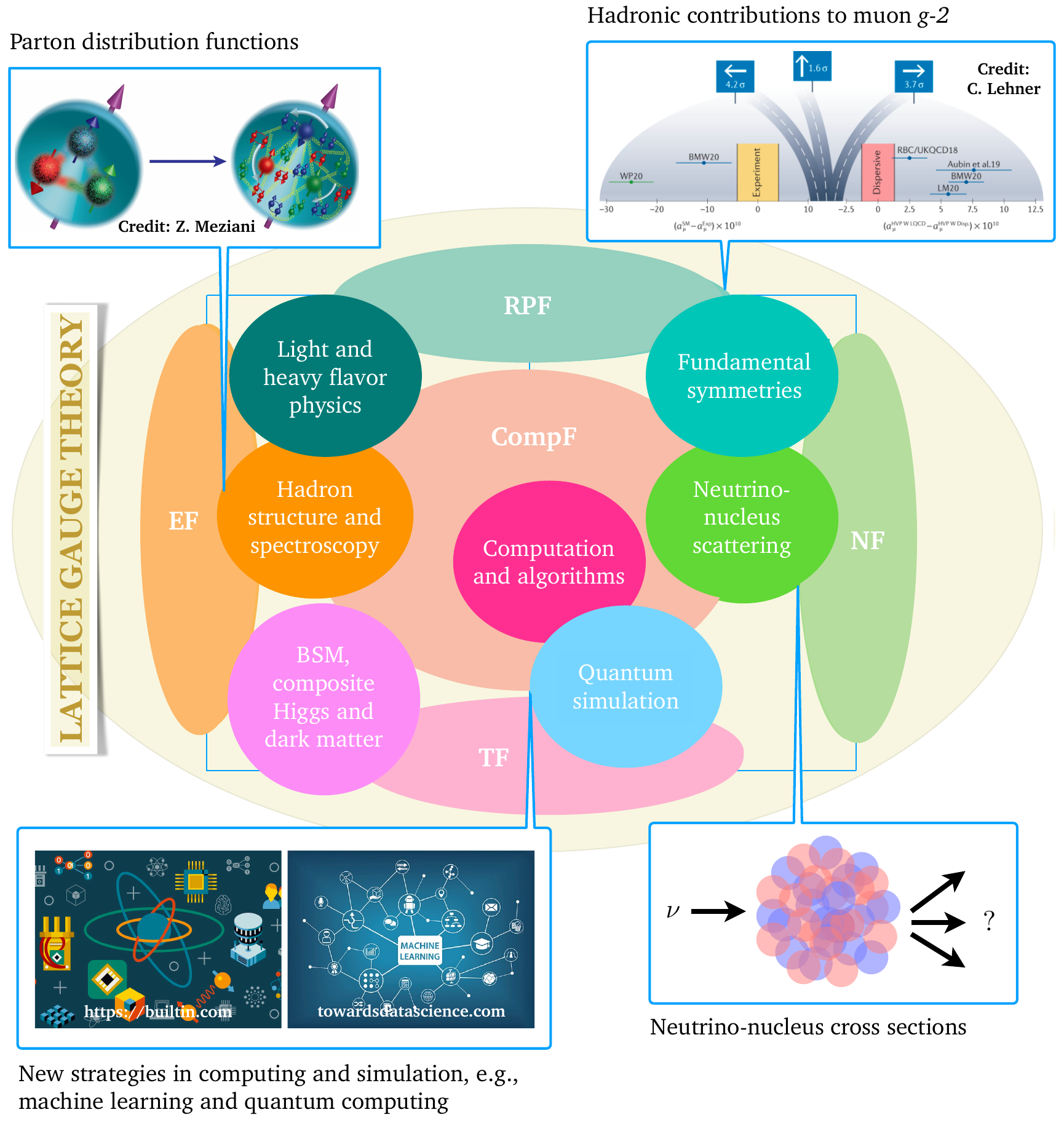}
\caption{The lattice-gauge-theory program touches on many areas of research in HEP as classified within the topical Frontiers of \emph{Snowmass2021} (TF: Theory Frontier, NF: Neutrino Frontier, RPF: Rare Processes and Precision Measurement Frontier, EF: Energy Frontier, and CompF: Computational Frontier). Areas with high impact and/or high levels of innovation along with select examples are highlighted in the figure.
}
\label{fig:graphic}
\end{figure}

The Particle Physics Community Planning Exercise, a.k.a.~``Snowmass'', provides an opportunity for the community to identify and document a scientific vision for the future of particle physics in the U.S. and its international partners~\cite{snowmass}. \emph{Snowmass2021} (the completion of which was delayed to 2022 due to the COVID pandemic) includes various scientific ``Frontiers'' such as Energy Frontier, Neutrino Physics Frontier, Rare Processes and Precision Measurement Frontier, Cosmic Frontier, Theory Frontier, Computational Frontier, and Community Engagement Frontier among others. Each Frontier consists of various topical working groups, each targeting specific research and development areas within the larger program, along with a number of liaisons who facilitated the coordination among frontiers with overlapping scientific agendas. The topical group in ``Lattice Gauge Theory'' (TF05) belongs to the Theory Frontier; nonetheless, given the broad application of lattice field theory in various scientific frontiers in HEP, as well as its close synergy with the computational research, it was well represented in four other Frontiers.

The lattice-gauge-theory community engaged in the planning exercise by submitting nearly 60 Letters of Interest to TF05 as primary or secondary category, and nearly 15 whitepapers with major focus on the role of lattice gauge theory in the HEP research and on new theory and computing approaches. A number of workshops and conferences were co-organized by the lattice gauge theorists within the Snowmass process on topics such as rare processes and precision measurements, heavy-flavor physics and CKM matrix elements, and neutrino-nucleus scattering. TF05 also organized dedicated sessions and talks at the (virtual) Snowmass Community Planning Meeting in October 2020, at the Theory Frontier Conference at the Kavli Institute for Theoretical Physics in Santa Barbara, CA in February 2022, and at the Snowmass Community Summer Study Workshop at the University of Washington, Seattle, WA in July 2022.

This report is an attempt to summarize the many conclusions arrived in the \emph{Snowmass2021} planning exercise by a vibrant community of lattice gauge theorists in particular, and high-energy physicists in general, on the critical role of the lattice-field-theory research on the experimental program in HEP, on the state-of-the-art computational paradigms, and on workforce and career development.\footnote{Various applications of lattice gauge theory in HEP and planned developments are additionally covered in Topical Reports within the Energy Frontier,
the Neutrino Frontier,
the Rare Processes and Precision Measurement Frontier,
and the Computational Frontier
of \emph{Snowmass2021}.} Substantial discussions in this report are drawn directly from a number of whitepapers that were co-solicited by the TF05 conveners and by the USQCD Collaboration~\cite{Colangelo:2022jxc, dEnterria:2022hzv, Blum:2022wsz, Ruso:2022qes, Alarcon:2022ero, Cirigliano:2022oqy, Constantinou:2022yye, Bulava:2022ovd, Batell:2022pzc, Catterall:2022qzs, Boyle:2022ncb, Boyda:2022nmh, Boyle:2022uba,
Bauer:2022hpo, USQCD:2022mmc,USQCD:2019hee, Cirigliano:2019jig}, along with a number other important whitepapers and Letters of Interests as referenced throughout the report. For further details, analysis, and references, the reader should consult these community-driven studies.

\section{Enabling precision quark- and lepton-flavor physics with lattice QCD}

\subsection{Muon anomalous magnetic moment and tau decay}
The current disagreement between experimental measurements \cite{Muong-2:2006rrc,Muong-2:2021ojo} and theoretical predictions \cite{Aoyama:2020ynm} for the anomalous magnetic moment of the muon $a_\mu$ (also known as ``muon $(g-2)$'') is one of the most significant outstanding tensions with the Standard Model, with a current statistical significance of 4.2$\sigma$ \cite{Colangelo:2022jxc}.  The experimental uncertainty will reduce significantly with additional data taking by the Fermilab Muon $(g-2)$ Experiment, with the potential for further reductions by future experiments at J-PARC \cite{Abe:2019thb}, PSI \cite{Aiba:2021bxe}, or at Fermilab \cite{Colangelo:2022jxc}. Further reductions in theory error are and will continue to be essential to bound, or potentially discover, the presence of new physics beyond the Standard Model in muon $(g-2)$.

The present theory error is dominated by uncertainties in two sets of contributions which depend on hadronic physics: hadronic vacuum polarization (HVP), and hadronic light-by-light scattering (HLbL).  Each of these contributions may be estimated using a ``data-driven'' approach, which uses dispersive analysis to relate the HVP and HLbL to experimental measurements of $e^+ e^-$ annihilation to hadrons and other measured properties of specific hadronic states, predominantly pions.  However, there are tensions between results from the BaBar and KLOE experiments which must be resolved by new experiments and further study, in order to reach the precision goal of below 0.3\% for the HVP contribution
by 2025. Since the magnitude of the HLbL contribution is smaller than HVP by more than a factor of 50, a 10\% precision goal in HLbL contribution will be more than commensurate with the HVP precision goal.

Lattice gauge theory provides an essential complement to the data-driven approach, allowing calculation of the HVP and HLbL contributions \emph{ab initio} without direct reliance on experimental inputs.  Lattice-QCD results are projected to reach similar precision levels to the data-driven approach by 2025, below 0.5\% for HVP and 10\% for HLbL.  Results from the two approaches may be combined to obtain a final theory result with the best possible precision, as has already been done for the HLbL contribution in Ref.~\cite{Aoyama:2020ynm}. HVP results can also be improved by such a combination, but this was not done in Ref.~\cite{Aoyama:2020ynm} as the lattice-QCD calculations are not yet precise enough compared to the data-driven results for the combination to be useful.  Reaching these precision goals will require improved statistics as well as good control over various sources of systematic uncertainty, including continuum extrapolation (which can be improved by calculation at smaller lattice spacings, although this requires a resource-intensive computational effort), strong isospin breaking (as most high-statistics lattice-QCD calculations are carried out at $m_u = m_d$, a small correction is needed), and infinite-volume extrapolation (which is complicated by long-distance contributions from two-pion states, although these may be studied independently by exclusive-channel calculations).  For HLbL, the primary systematic effect of concern has to do with volume dependence in treatment of the photon IR modes. At this point, results from independent groups using different methods for the photon are in good agreement, and the overall lattice-QCD results are in agreement with the data-driven estimate.

There is some tension between the most precise current lattice-QCD determination of HVP by \cite{Borsanyi:2020mff} and the world-average data-driven result from Ref.~\cite{Aoyama:2020ynm}, with the lattice-QCD result agreeing more closely with the experimental result for $a_\mu$.  It will be crucial for more lattice collaborations to perform independent calculations of HVP with similar precision to clarify the situation.

For comparisons between lattice-QCD results, the use of Euclidean-time windows which restrict the analysis to an intermediate time range are extremely useful. This removes the largest contributions from lattice discretization (at short times) and finite-volume effects (at long times) and allows for the most precise comparisons between independent calculations.  There is a mild tension between some current lattice-QCD results for intermediate-time window for HVP, with further results forthcoming from additional groups, see Ref.~\cite{Colangelo:2022jxc} for a summary of the current situation, with more recent lattice-QCD results for the intermediate window in Ref.~\cite{Wang:2022lkq,Aubin:2022hgm,Ce:2022kxy,Alexandrou:2022amy}.  Aside from facilitating comparisons between lattice-QCD results, the use of time-window results may also provide an ideal way to combine lattice-QCD and data-driven results over the complementary time/energy ranges where they are most precise~\cite{RBC:2018dos}.

On a longer time horizon, it will continue to be important for lattice-QCD calculations to pursue higher precision with uncertainties below the 2025 uncertainty goals, in order to fully leverage the Fermilab (g-2) result as well as to prepare for future muon $(g-2)$ experiments.  This will be achieved through a combination of increased computational power which should allow for calculations at even larger physical volumes and smaller lattice spacings, and improved methods and algorithms for gauge-ensemble generation and measurement with reduced statistical noise.

Aside from muon $(g-2)$, another leptonic process for which lattice QCD can provide crucial input is the hadronic decay of $\tau$ leptons.  The CKM matrix element $|V_{us}|$ can be determined via study of inclusive $\tau$ decays~\cite{RBC:2018uyk,Aoki:2021kgd}.  Lattice study of the inclusive process is done using a calculational setup which is very similar to the muon $(g-2)$ HVP.  Study of $\tau$ decays can also be useful in searching for possible new physics~\cite{Cirigliano:2021yto}, where in addition to the inclusive processes it is useful to study exclusive processes such as $\tau \rightarrow (\pi/K) \nu$ and $\tau \rightarrow (\pi/K/\eta) \pi \nu$.  The former processes rely on lattice-QCD determinations of $f_{\pi,K}$, while direct calculations of form factors on the lattice for the latter processes could help to reduce theoretical uncertainty.  Finally, there is also a connection between $\tau$ decays and muon $(g-2)$, where $\tau \rightarrow \pi \pi \nu$ can be related to $e^+ e^- \rightarrow \pi^+ \pi^-$ by way of lattice-QCD calculations that correct for the different isospin of the two-pion states~\cite{Bruno:2018ono}.

\subsection{Quark masses and strong coupling constant}
Quark masses, beyond their inherent value as fundamental parameters of the Standard Model, are also highly interesting in certain physics contexts.  The up-quark mass is especially relevant in the context of the strong CP problem, where $m_u = 0$ has been proposed as a potential solution \cite{tHooft:1976rip,Hook:2018dlk}; however, lattice-QCD simulations \cite{Aoki:2021kgd} have now determined $m_u \neq 0$ to very high confidence. The bottom and charm quark masses also determine the associated Yukawa couplings of the Higgs boson, and so are essential inputs for precision study of $H \rightarrow bb$ and $H \rightarrow cc$ decay modes at current and future colliders \cite{Lepage:2014fla}.  Current lattice-QCD efforts have led to percent-level precision on both $m_b$ and $m_c$, which is sufficient to meet the needs of even future proposed Higgs factories; see Refs.~\cite{Boyle:2022ncb,USQCD:2022mmc} for further details.

The strong coupling constant at the $Z$-pole, $\alpha_s(M_Z)$, is also an important input for high-precision determination of the Higgs boson couplings, as well as a variety of other theoretical predictions such as the hadronic width of the $Z$ boson.  Although the current determinations of $\alpha_s$ are likely sufficient for the LHC Higgs program, future studies at ``Higgs factory'' lepton colliders may require further improvements in $\alpha_s(M_Z)$ to make full use of the experimental reach~\cite{USQCD:2022mmc}. The current lattice world average determination for $\alpha_s(M_Z)$ reported by the Flavor Lattice Averaging Group (FLAG) has a precision of about 0.7\%~\cite{Aoki:2021kgd}.  Improvement of this precision in the future is limited by systematic errors, primarily associated with perturbative matching at relatively low energy scales. Further development of ``step-scaling'' methods may provide a way to obtain lattice results at higher energies and obtain $\alpha_s$ at a level of precision below 0.5\%~\cite{dEnterria:2022hzv}.

\subsection{Light-quark flavor physics}
Light-quark flavor physics is a notable example of the importance of lattice-QCD research at the intensity frontier, impacting the search for new physics at very high energies.  Deriving a theoretical prediction of decay and mixing of pions and $K$ mesons in  the Standard Model from the quark-level CKM theory 
has been one of the most active areas both in experimental and lattice-QCD communities.
While  the results of calculations of simple physical quantities have been improved in accuracy, more involved calculations such as those involving multi-hadron states, which were previously very difficult, have become possible with the development of computational methods, techniques, software, and hardware.

From experimental results on leptonic and semileptonic decays of pions and kaons, precise estimates of the CKM matrix elements $|V_{us}|$ and $|V_{ud}|$ can be obtained, providing important constraints on new physics through tests of the unitarity of the CKM matrix.  This requires precise theoretical determinations for decay constants $f_K/f_\pi$ and the semileptonic vector form factor $f_+(0)$. After decades of efforts by lattice-QCD researchers, these quantities have been determined with a very high precision of about 0.2\%~\cite{Aoki:2021kgd}.  At this level of precision, it is essential to maintain control over systematic effects such as corrections due to quantum electrodynamics (QED)~\cite{DiCarlo:2019thl}.  Surprisingly, the FLAG world average currently reported is in approximately 3 $\sigma$ tension with CKM unitarity, warranting further cross checks and future improvements.  

The direct (parametrized by $\epsilon’$) and indirect CP violation ($\epsilon$) in $K_L\to \pi\pi$ 
decay involving all three generations of quarks have great sensitivity to possible new sources of CP violation and, at the same time, need significant advances in lattice-QCD calculation~\cite{Blum:2022wsz}.
The progress of the current state-of-the-art lattice-QCD calculations of $\epsilon’ / \epsilon$ have achieved about 40\% precision~\cite{Abbott:2020hxn,Blum:2022wsz} and is consistent with experimental results from NA28 and KTeV.
The calculation relies on chiral lattice quarks (domain wall fermions, DWF),  the theoretical treatment of multi-hadron processes in Euclidean finite-volume spacetime (Lellouch-L\"uscher formalism), and the use of operator renormalization and matching between lattice and continuum  (RI/SMOM nonperturbative renormalization).  The final $\pi\pi$ state of the kaon decay on a finite-volume lattice is prepared in two complementary 
ways: as a lowest-energy state of two pions obeying G-parity boundary conditions, and as an excited state with periodic boundary conditions. 

The very small mass difference between $K_L$ and $K_S$, $\Delta M_K=3.482(5)\times 10^{-12}$ MeV, gives a possibility to constrain new physics at very high scales, $O(1000)$ TeV. A Standard-Model calculation of $\Delta M_K$ with lattice QCD implementing the important charm quark loop and GIM mechanism 
was reported with 40\% precision~\cite{Bai:2014cva,Wang:2021twm,Blum:2022wsz} together with the closely related long-distance contribution to $\epsilon$.  This research identified and removed the unwanted contribution from $K, \pi, \pi\pi, \eta$ intermediate states between two electroweak operators, an artifact of Euclidean spacetime which appears in many other important processes involving multi-hadron intermediate states.

Future plans for $\epsilon’ / \epsilon$ calculation include improvements of the discretization error using  fine lattice spacings (with $\sim 3, 4, 5.5$ GeV lattice cutoffs), incorporating virtual charm quarks, theoretical and computational studies of the effect of isospin breaking, which is enhanced by the $\Delta I = 1/2$ rule,  and higher-order perturbative-QCD calculations for the four-point quark operators in the effective electroweak Hamiltonian. The next decade should see a reduction of the current uncertainty by a factor of about 4, with estimated computational cost required being approximately 2000 exaflop-hours.  The lattice-QCD community further plans to extend the light-quark studies to 
rare Kaon decays and to inclusion of electromagnetism in various light-flavor processes to meet the required precision goals in various searches for deviations from the Standard Model.

\subsection{Heavy-quark flavor physics}
Electroweak decays of bottom and charm quarks provide fertile experimental ground for rigorous tests of the Standard Model, with the potential to uncover new physics.  There have been persistent tensions between theory and experiment for several years, most prominently in decays involving $B$ mesons~\cite{USQCD:2019hyg,Gambino:2020jvv,Lenz:2021bkv}.  Although no single deviation is significant enough to provide unambiguous evidence of new physics, anticipated improvements from ongoing and future experiments such as Belle II, LHCb, ATLAS, CMS, and BES III may lead to the discovery of new physics in these channels.  However, the possibility of discovering new physics will require simultaneous improvements in theoretical precision.  Lattice QCD has played a central role in theory predictions related to a wide range of electroweak decay processes involving charm and bottom quarks, in several cases at levels of precision smaller than current experimental uncertainties.  A detailed review of lattice QCD and heavy-flavor decays can be found in a recent Snowmass whitepaper~\cite{Boyle:2022uba}.

Systematic improvement of existing calculations, as well as development of altogether new methods to access processes which have not yet been explored on the lattice, will be crucial for understanding the nature of these anomalies in the years to come.  For processes such as $B \rightarrow \mu^+ \mu^-$ or $D \rightarrow K \ell \nu$, theory errors are already well under control, and future improvements will rely on tackling small sources of systematic uncertainty such as strong-isospin breaking and QED effects. Improvements of the precision of lattice calculations for $B$-meson semileptonic decays are still required to match the precision of Belle II; here, smaller lattice spacings and high statistics can have a large impact.  Other decays with current anomalies such as $B \rightarrow K^* \ell^+ \ell^-$ with unstable hadrons in the final state will require further work, but should yield useful lattice-QCD results in the near future.  The required methodology to treat $1 \rightarrow 2$ hadronic processes in lattice QCD has been fully developed as demonstrated by the results for $K \rightarrow \pi \pi$ (discussed above), but the presence of multiple final states due to the kinematics of heavy-meson decays makes such calculations significantly more challenging.  Finally, some channels for understanding $B$ anomalies require new methods at the forefront of theoretical understanding.  For example, tensions between inclusive and exclusive determinations of CKM matrix elements can be probed directly if lattice-QCD calculations of inclusive decay processes could be done.  Such a calculation is equivalent to extraction of the (Minkowski-spacetime) spectral density from (Euclidean-spacetime) correlation functions, which is theoretically difficult, but the new ideas proposed in recent years~\cite{Hansen:2017mnd,Hashimoto:2017wqo,Hansen:2019idp,Gambino:2020crt,Maechler:2021kax,DeGrand:2022lmc} provide a path forward.

Although the $B$ anomalies and other potential searches for new physics provide an interesting and timely motivation for the study of heavy-quark decays, these decay processes are also of critical importance for determination of fundamental parameters of the Standard Model: the CKM matrix elements. Precise calculations of form factors and decay constants from lattice QCD are essential in using experimental results to extract the CKM matrix elements.  A wide array of heavy-quark processes are relevant in this context, including neutral $B$ meson mixing and lifetimes, leptonic and semileptonic decays of $B$ and $D$ mesons, radiative decays (including real photons in the final state), and decays of baryons containing heavy quarks.  As in the specific examples given above, future progress will rely on a mixture of refinement of existing calculations to higher precision, and new theoretical developments to enable calculations of more complex decay processes from lattice QCD.  

\section{Uncovering new-physics signals in nucleons and nuclei with lattice QCD
\label{sec:FS}}

\subsection{Neutrino-nucleus scattering for neutrino phenomenology
\label{sec:nu-nucleus}}   
Neutrino oscillation experiments have provided direct evidence that neutrinos have nonzero masses arising from new BSM interactions. Future accelerator neutrino experiments such as Deep Underground Neutrino Experiment (DUNE) and Hyper-Kamiokande aim to determine the mass and mixing parameters governing neutrino oscillations to unprecedented precision, shed light on the neutrino mass hierarchy and the presence of CP violation in the lepton sector, and search for potential signals of new physics. 
The analyses of these experiments require as input the incident neutrino energy, which is \emph{a priori} unknown. Precise predictions of neutrino-nucleus cross sections are, therefore, necessary input to neutrino event generators that are used to reconstruct the neutrino energy, and to estimate the expected event rates~\cite{Campbell:2022qmc}. As the experimentally-common target media are nuclear isotopes such as carbon, oxygen, and argon, deep knowledge of the underlying nuclear physics of response of nuclei to electroweak probes is crucial. Unfortunately, such knowledge is currently incomplete for both nucleonic and nuclear-level quantities that enter the description of relevant cross sections~\cite{Ruso:2022qes}. Only a concerted effort in combining methods such as lattice QCD, nuclear effective theories, phenomenological models, and \emph{ab initio} nuclear many-body theory can ensure that reliable cross sections are provided to the event-generator community. Lattice QCD will play a crucial, and in some instances complementary, role in this theoretical endeavor. 

Drawing conclusions from recent community whitepapers~\cite{Ruso:2022qes,Kronfeld:2019nfb}, one can enumerate the essential quantities in this program that can be accessed via lattice QCD. Depending on the neutrino energy, $E_\nu$, several different reaction channels can contribute to the final-state event rates, each demanding certain nonperturbative quantities to be evaluated:
\begin{itemize}
\item[--] When $E_{\nu} \lesssim 100$ MeV, the main interaction channels are ``coherent elastic neutrino-nucleus scattering" (CEvNS) and exclusive inelastic processes involving nuclear excited states. The nucleon electric and magnetic form factors that contribute coherently to CEvNS cross sections are known precisely from electron-scattering experiments~\cite{Ankowski:2022thw}, and the largest uncertainties arise from nuclear structure functions including neutron distributions; see~\cite{Ruso:2022qes} for more details. Lattice-QCD calculations of nucleon axial form factors and of multi-nucleon correlations and currents could reduce other subdominant but significant uncertainties in low-energy cross section calculations.

\item[--] When $E_{\nu} \sim 0.1 - 1$ GeV, the dominant reaction mechanism is “quasielastic scattering”, in which the nucleus transitions to another state but the final state includes a lepton and a nucleon that can be detected in experiment. Both vector and axial-vector currents make significant contributions to quasielastic scattering. Nucleon axial and induced pseudoscalar form factors are, therefore, crucial ingredients to calculations of charged- and neutral-current neutrino-nucleus scattering cross sections within nuclear many-body models and neutrino event generators. Lattice-QCD calculations are already reaching competitive precision with experimental determinations of axial form factors~\cite{Shintani:2018ozy,RQCD:2019jai,Alexandrou:2020okk,Meyer:2021vfq,Ishikawa:2021eut,Schulz:2021kwz}.

\item[--] When $1$--$3~{\rm GeV} \lesssim E_\nu \lesssim 3~{\rm GeV}$, reaction mechanisms involving resonance production dominate the cross section and interactions with correlated pairs of nucleons also contribute significantly.
This is called the ``resonance--production region". Precise determinations of $\Delta$, Roper, and other $N^*$ resonance properties and transition form factors from lattice QCD would inform models of resonant neutrino scattering by providing results that will be complementary to experimental pion-nucleon and neutrino scattering data. Lattice QCD has come a long way over the past decade to enable accessing the physics of resonances and excited spectrum of QCD, as noted in Sec.~\ref{sec:spec-struc}. Preliminary calculations of nucleon-pion scattering and transition form factors have been reported~\cite{Andersen:2017una,Silvi:2021uya,Barca:2021iak,Bulava:2022vpq}, and realistic calculations of the relevant transitions in the neutrino-nucleus scattering in this kinematic range could be in reach.

\item[--] When $E_\nu \sim 3-5~{\rm GeV}$, multiple pions and a variety of $N^*$ resonances can be produced in the final state, a process which is called ``shallow inelastic scattering". It is presently computationally infeasible to include the multitude of $N\pi$, $N\pi\pi$, $N\pi\pi\pi$, and other multi-hadron scattering states in the calculations. Besides the complexity of correlation functions that need to be constructed and evaluated, a dense multi-particle spectrum in the finite volume of lattice-QCD calculations makes controlling uncertainties associated with excited-state effects challenging. Furthermore, the existing formalisms that connect scattering and resonant properties to the lattice-QCD output are complex, and the generalization of such formalisms, or EFT-based matchings, for more than three-hadron systems are not well developed. Formal progress in this area is expected over the next decade. On the other hand, lattice-QCD calculations of nucleons' hadron tensors could give access to inclusive cross sections when decompositions into exclusive channels are infeasible. First efforts at computing this quantity using lattice QCD have been reported~\cite{Liang:2019frk,Fukaya:2020wpp}, but further progress is challenged by the need to solve an ill-posed inverse problem to relate the computationally accessible Euclidean hadron tensor to its experimentally relevant Minkowski counterpart, see Sec.~\ref{sec:pdf}. Advancing solutions to this problem will likely constitute an active area of research in the coming years, and could bridge the existing gap between the shallow- and deep-inelastic regions in neutrino-nucleus scattering.

\item[--] When $5~{\rm GeV} \lesssim E_\nu$, one enters the ``deep inelastic region", where the underlying physics description is simplified by the factorization of hard scattering amplitudes calculable in perturbative QCD and nonperturbative PDFs. As discussed in Sec.~\ref{sec:pdf}, parton distribution functions are not directly accessible in lattice QCD since they are defined via a Minkowski light-cone matrix element of partonic-level operators in the nucleon while lattice-QCD computations are performed in Euclidean spacetime. PDFs can, however, be accessed indirectly via resorting to a Mellin-moment expansion of the PDFs, where lattice QCD can provide the first few moments by computing matrix elements of local operators that are insensitive to the time signature of spacetime. PDFs can also be accessed via quasi-PDFs (or related quantities that are Euclidean counterparts of the Minkowski PDFs) that are evaluated in the large-momentum frame of the nucleon and can be matched to the true PDFs in the limit of infinite momentum perturbatively. Significant progress has been reported in recent years in this problem, see Sec.~\ref{sec:pdf}, and it is expected that over the next few years, reliable lattice-QCD determinations of nucleon isovector unpolarized and polarized PDFs at intermediate Bjorken-$x$ parameter will provide important inputs for neutrino-nucleus scattering cross sections in the deep-inelastic scattering region.

\end{itemize}

Not only can lattice QCD provide critical single-nucleon input, it can also crucially inform the nuclear many-body calculations of experimentally relevant nuclei by illuminating and quantifying nuclear effects relevant for neutrino interactions. Nuclear EFTs characterize systematically single-, two-, and higher-body effects in form factors, transition amplitudes, and structure functions in terms of a number of low-energy constants in interactions and currents, whose values can be constrained by matching to lattice-QCD calculations in light nuclei (when experimental input is limited). Proof-of-principle studies of nuclear effects, including: in electroweak transition amplitudes of light nuclei~\cite{Beane:2015yha,Savage:2016kon,Shanahan:2017bgi,Tiburzi:2017iux,Parreno:2021ovq,Davoudi:2020ngi}; in (the zero-momentum-transfer limit of) their vector, axial, scalar, and tensor form-factors~\cite{Chang:2017eiq}; and in their partonic momentum fractions~\cite{Detmold:2020snb} have emerged in recent years, and improved calculations with better-controlled uncertainties are planned over the next decade. Constraining the corrections dominantly arising from two- and higher-body current interactions with correlated pairs of nucleons using lattice QCD will be another important contribution to the neutrino-nucleus scattering program.

For both nucleon and multi-nucleon calculations, an important remaining challenge is to control and isolate contaminating excited-state contributions at earlier Euclidean times in the face of an exponentially degrading signal-to-noise in nucleonic correlation functions at later times. The development and application of variational techniques in accessing spectral and transition amplitudes in lattice QCD~\cite{Francis:2018qch,Horz:2020zvv,Amarasinghe:2021lqa} will continue in the coming years to systematically address these effects. Notably,  constraints on the resonant transition amplitudes in the resonant-production region enable the quantification of the excited-state effects due to these resonances in the elastic and quasielastic regions. Hence, lattice-QCD calculations of various quantities stated above are complementary to advancing this program.

\subsection{Electric dipole moments for probing CP violation}
Successful calculations of the contributions of CP-violating operators to the neutron, proton, and nuclear electric dipole moments (generically called nEDM here) using lattice QCD are necessary to understand and quantify the fundamental nature of CP violation, a promising way to elucidate new physics beyond the TeV energy scale, including the mechanism for baryogenesis. Lattice QCD can provide the matrix elements of quark- and gluon-level operators such as the vacuum angle $\theta$, quark's (chromo-) EDM, Weinberg three-gluon and four-quark operators, etc., within hadronic states. The community envisions~\cite{Alarcon:2022ero} deriving the contribution of physics beyond the Standard Model to nEDM by combining these lattice-QCD calculations with their coupling strengths at the hadronic scale obtained using EFT methods. To fully utilize the predictive power of planned experimental limits realizable in the next decade, e.g., $10^{-28}$~$e\cdot$cm for neutron EDM~\cite{NEDMsummary1,NEDMsummary2} and $10^{-29}$~$e\cdot$cm for proton EDM from the storage-ring experiments~\cite{Alexander:2022rmq}, a minimal target uncertainty of ${}\sim25\%$ is required for the CP-violating part of the matrix elements of the operators with mass dimensions up to 6~\cite{Chien:2015xha}, whereas currently some of them are known to only an order of magnitude.

In the past decade, a great deal of progress has been made in the calculations of the EDM of the nucleon. Computations include those for the dimension-4 vacuum angle $\theta$ (the strong CP problem), the quark EDM and quark chromo-EDM operators at dimension-5, and the dimension-6 Weinberg three-gluon operator. New techniques and formalism to correctly extract the CP-violating part of their matrix elements within the nucleon ground state, i.e., the form factor $F_3(Q^2)$, are now developed~\cite{Abramczyk:2017oxr,Bhattacharya:2018bkd,Shindler:2021bcx}. Despite the progress, the contribution of the $\theta$-term to nEDM remains unresolved, primarily due to the smallness of the coupling between the gluonic topological charge and spin dynamics of the nucleon, which vanishes in the chiral limit. Furthermore, possible lattice artifacts such as the contributions of light $N\pi$ multi-hadron excited states  can be large and are not yet fully resolved. Currently, the statistical uncertainties from the Monte-Carlo simulations at the physical light quark masses, as well as the systematics of the chiral extrapolations from larger masses, remain leading sources of uncertainties~\cite{Bhattacharya:2021lol}. The situation is much better for the contribution of the quark-EDM operator, which is given by the flavor-diagonal tensor charges, and results with a total uncertainty of  ${} < 5\%$ are available~\cite{Aoki:2021kgd}. The quark chromo-EDM operator and the Weinberg three-gluon operator have complicated renormalization and mixing structures~\cite{Bhattacharya:2015rsa,Mereghetti:2021nkt}, and as of now, only preliminary calculations exist~\cite{Shindler:2021bcx,Bhattacharya:2022whc}. The CP-violating four-quark operators have not yet been investigated using first-principles techniques like lattice QCD.

Interpretation of experimental results on atoms and nucleons also needs nuclear- and atomic-physics calculations using as inputs the CP-violating electron-nucleon and nucleon-nucleon couplings, the important long-distance part of the latter being mediated by the CP-violating pion-nucleon coupling.  Of these, calculations in the last decade have provided the matrix elements necessary for the leading CP-violating electron-nucleon couplings at a few to 10\% level~\cite{Aoki:2021kgd}. Methods have also been proposed for the calculation of the CP-violating pion-nucleon couplings~\cite{Shindler:2021bcx,Bhattacharya:2022whc}, but no results are available yet. These calculations will continue to progress and mature over the next decade to support a vibrant EDM experimental program~\cite{Blum:2022cie}.

\subsection{Baryon and lepton number/flavor nonconservation}

\subsubsection{Lepton-number nonconservation and neutrinoless double-$\beta$ decay}
Searches for lepton-number-violating interactions leading to neutrinoless double-$\beta$  ($0\nu\beta\beta$) decay probe fundamental questions about the nature of neutrino masses and will provide complementary insights to neutrino-oscillation experiments. The search for a $0\nu\beta\beta$ decay will intensify with the forthcoming commissioning of ton-scale experiments in the U.S. and worldwide~\cite{Bilenky:2014uka, DellOro:2016tmg, Dolinski:2019nrj}. A potential discovery will have profound implication for our understanding of neutrinos, i.e., whether they are of a Dirac or Majorana type, and will elucidate the existence of one or more plausible lepton-number-violating (LNV) scenarios at high-energy scales. Nonetheless, given that such a decay is only expected for a handful of medium- and high-mass atomic isotopes, deciphering any potential decay signal and tracing it back to the underlying LNV mechanism demands a multi-faceted and concerted theoretical campaign that involves both high-energy and nuclear physicists. As argued in a recent Snowmass whitepaper~\cite{Cirigliano:2022oqy}, an overarching goal of this program is to compute $0\nu\beta\beta$ rates with minimal model dependence and quantifiable theoretical uncertainties by advancing progress in particle and nuclear EFTs, lattice QCD, and nuclear few- and many-body \emph{ab initio} methods. 

By matching BSM models of $0\nu\beta\beta$ to hadronic EFTs, a systematic expansion of the $0\nu\beta\beta$ rates is possible in the few-nucleon sector~\cite{Prezeau:2003xn,Graesser:2016bpz,Cirigliano:2018yza,Cirigliano:2018hja,Cirigliano:2019vdj,Dekens:2020ttz}, but these EFTs need to be complemented with values for low-energy constants (LECs). With the lack of experimental input, lattice QCD, which computes the relevant matrix elements directly, is the only way to determine these LECs with quantifiable uncertainties. Currently, the first computational targets for lattice QCD are the $nn \to pp$ matrix elements in various LNV scenarios. These matrix elements can be classified into two categories: i) matrix elements of local quark-lepton-level operators that have resulted from LNV scenarios involving a high mass which has been integrated out down to the QCD scale, and ii) matrix elements of nonlocal quark-lepton-level operators that involve the propagation of a Majorana neutrino of either low mass (referred to as the minimal extension of the Standard Model) or sterile neutrinos with a mass that cannot be integrated out at the QCD scale. 
 
 The necessary lattice-QCD calculations for both classes proceed in three stages. First, two-nucleon spectra and elastic scattering amplitudes need to be computed and constrained, then the QCD two-nucleon matrix elements in the relevant scenario are determined and computed, and finally the physical infinite-volume, Minkowski-space transition amplitudes are extracted, via direct or indirect mapping to the EFT descriptions that feed into the nuclear many-body calculations. With the latter step being partly developed for various matrix-element classes in recent years~\cite{Lellouch:2000pv,Briceno:2015tza,Davoudi:2020gxs,Feng:2020nqj} and continuing to be advanced, the challenge in the coming years would be to achieve accurate and precise determination of the two-nucleon spectra and matrix elements. Lattice-QCD calculations of two-nucleon spectra and scattering are necessary for ensuring that operators that couple well to the physical two-nucleon systems are identified for the use in $nn \to pp$ processes, systematic uncertainties are identified and sufficiently controlled for NN observables, and infinite-volume transition amplitudes can be extracted from finite-volume matrix elements, a process that requires access to NN finite-volume energies and energy-dependence of the NN scattering amplitudes near transition energies. The two-nucleon spectroscopy even at unphysically large quark masses has proven challenging given a severe signal-to-noise degradation issue and a dense excited-state spectrum as discussed in Sec.~\ref{sec:spec-struc}, but the use of variational techniques which have begun in the NN sector~\cite{Francis:2018qch,Horz:2020zvv,Amarasinghe:2021lqa} and increased computing resources in the exascale era may resolve the situation in the near term. First calculations in simpler pionic systems are complete~\cite{Tuo:2019bue, Detmold:2020jqv, Nicholson:2018mwc, Detmold:2022jwu}, and the required precision on the NN spectra and matrix elements from lattice QCD to impact the nuclear many-body calculations of experimentally relevant nuclei are identified in a minimal extension of the Standard Model~\cite{Davoudi:2021noh}. Such studies will continue to guide the computational effort in the coming years.

The community has further identified the next stages of the program, once the first and yet challenging goals stated above are achieved~\cite{Cirigliano:2022oqy}. These include the evaluation of a range of matrix elements (for example those involving both pions and nucleons) relevant for constraining contributions at higher orders in the pertinent hadronic EFTs, hence allowing for tests of the EFT power counting, as well as evaluating the $nn \to pp$ matrix elements in light nuclei to quantify multi-nucleon effects,  guiding the EFT descriptions.

\subsubsection{
Baryon-number nonconservation, proton decay, and $n-\bar{n}$ oscillation}
There is a long history of experimental searches for baryon-number-violating interactions that lead to proton decay. Current experimental bounds from Super-Kamiokande will be significantly improved by upcoming searches at DUNE and Hyper-Kamiokande. To constrain BSM theories that predict observable proton decay, matrix elements of three-quark operators between experimentally relevant initial and final states are needed. Several lattice-QCD calculations of these matrix elements have been computed using a proton initial state~\cite{CP-PACS:2004wqk,Aoki:2008ku, Aoki:2013yxa,Aoki:2017puj, Yoo:2021gql}, including recent calculations with approximately physical quark masses. A challenge for future studies will be to consider nuclear effects on proton-decay matrix elements and provide further QCD input to nuclear many-body calculations of proton decay in nuclei.

Complementary to the searches for baryon-number (B) violation in proton decay and lepton-number (L) violation in $0\nu\beta\beta$ decay, neutron-antineutron ($n-\bar n$) oscillations probe the violation of baryon number by 2 units ($\Delta B = 2$), and can directly probe breaking of the Standard Model conserved quantity B-L (unlike proton decay, which conserves B-L.) Constraints on the oscillation time from experiments can test several low-scale baryogenesis scenarios~\cite{Mohapatra:2009wp,Babu:2013yww,Phillips:2014fgb}, and are therefore extremely valuable. The best limits on the oscillation time come from oscillation detection in nuclei, and are expected to be further improved in DUNE. Constraints also arise from cold neutron-beam experiments~\cite{Baldo-Ceolin:1994hzw} and more sensitive searches will be performed at a proposed experiment at the European Spallation Source~\cite{Addazi:2020nlz}. These measurements are theoretically clean and can be related to constraints on the BSM physics directly using lattice-QCD results. In a Standard Model EFT, the $n-\bar n$ oscillations proceed via a set of six-quark operators whose matrix elements can be evaluated in the pertinent states using lattice QCD. If the initial and final states are set to the neutron and its antiparticle, the results can directly apply to the neutron-beam experiments. Such results have emerged in recent years with relatively large statistical uncertainties but reasonably well-controlled systematic uncertainties~\cite{Rinaldi:2018osy}. They point to up to an order of magnitude enhancement in the expected rate compared with quark-model estimates, hence improving the reach of current and future experiments into the BSM parameter space. While these calculations will improve in precision in the coming years, other feasible studies such as those involving higher-dimensional six-quark operators with electromagnetic current insertions can be pursued---a scenario that is motivated by candidate models of (B-L)-violating $n-\bar n$ oscillations not suppressed by the presence of magnetic fields~\cite{Gardner:2017szu}. 

If nucleon conversion occurs in nuclear media, the obtained rates in vacuum can only be contrasted with experiment upon an EFT or model matching to nuclear many-body calculations. As argued in a recent community whitepaper~\cite{Cirigliano:2019jig}, challenging calculations for lattice-QCD physicists will be the matrix elements of the same operators in light nuclei to inform the EFT analysis. The challenge, besides a severe exponential signal-to-noise degradation and enhanced excited-state contamination, is in the production of multiple final-state hadrons, which complicates translating the Euclidean matrix elements to physical oscillation amplitudes. Nonetheless, given the progress in multi-hadron physics from lattice QCD as reported in Sec.~\ref{sec:spec-struc}, and motivated by the experimental searches using deuterium~\cite{SNO:2017pha} that are augmented by a recently-developed EFT description~\cite{Oosterhof:2019dlo,Haidenbauer:2019fyd}, the lattice-QCD community is set to explore possibilities for calculations of six-quark matrix elements in two-nucleon systems that would provide unknown EFT inputs, and would inform power-counting choices, in the upcoming years.
 
\subsubsection{Lepton-flavor nonconservation and muon to electron conversion}
While the discovery of neutrino oscillations establishes the violation of lepton-flavor conservation in Nature, in minimal extensions of the Standard Model that include only neutrino masses and mixing, branching ratios for lepton-flavor-violating (LFV) decays of charged leptons are small ($<10^{-54}$)~\cite{Chang:1980ey, Kuo:1980ew}. Processes such as $\mu \to e$, $\mu \to e \gamma$, and $\mu  \to eee$ in nuclei are expected to probe other LFV mechanisms at scales up to $10^3$ TeV in the current and future experiments, with sensitivities to branching ratios in the range $10^{-13}-10^{-16}$. Great enhancements are expected, in particular, in next-generation experiments such as Mu2e at Fermilab and COMET and DeeMe at J-PARC, as discussed in a recent Snowmass whitepaper~\cite{Lee:2022moh}. The success of future searches of $\mu \to e$ conversion for the underlying LFV mechanisms depends upon accurate matrix elements of the associated operators in relevant nuclei such as aluminum and carbon. Again lattice QCD enters by providing the nonperturbative matrix elements in the nucleon for two sets of operators at dimension six: i) a range of LFV two-quark--two-lepton operators, and ii) quark bilinears coupled to an electromagnetic field which in turn converts the muon flying by to an electron. Lattice QCD can also compute the matrix elements in light nuclei to inform the EFTs that are used in nuclear many-body calculations to predict conversion rates in the isotopes of relevance in experiment.

The needed coupling of the external probe (quark density) to single nucleons are in turn a set of nucleon form factors (scalar, vector, axial, tensor, and pseudoscalar) where the relevant momentum transfer to nucleon is set by the mass of the muon. While controlled few-percent precision for the zero-recoil limit of the form factors, or namely charges, have been achieved with the lattice-QCD technique, calculations of various form factors will still need to be completed in the coming years with a full account of uncertainties~\cite{Gupta:2018lvp, Bali:2018qus, Ren:2014vea, Yang:2015uis, Durr:2015dna, Bali:2016lvx, Abdel-Rehim:2016won,Jang:2019jkn,Hasan:2017wwt,Park:2021ypf,Meyer:2022mix,Aoki:2021kgd}. 

As argued in a recent whitepaper~\cite{Cirigliano:2019jig}, a key challenge is reaching the low-momentum-transfer region of the form factors by enlarging the lattice volume or the use of appropriate boundary conditions. Furthermore, for the proposed LFV Higgs-mediated processes~\cite{Crivellin:2014cta, Crivellin:2017rmk}, a set of scalar form factors as well as the matrix elements of gluonic operators are needed. Additionally, for spin-dependent mechanisms, other tensor structures, especially the flavor-diagonal axial form factor, become relevant~\cite{Cirigliano:2017azj}. These are computationally more challenging, generally requiring higher statistics, but recent developments have allowed quantities of this type to be reliably estimated with lattice QCD~\cite{Aoki:2021kgd}. An even more challenging endeavor in the coming years is constraining the form factors in light nuclei in an attempt to quantify the nuclear modifications to the single-nucleon responses. Preliminary lattice-QCD calculations for nuclei with atomic number $A<4$ at large values of the quark masses point to large ($\sim 10\%$) nuclear modification in the scalar matrix element at zero recoil~\cite{Chang:2017eiq}, which may become significant in larger nuclei. This motivates the need for robust lattice-QCD constraints in light nuclei at the physical values of the quark masses.

\subsection{Precision $\beta$ decay for searches of new physics}
Besides the avenues listed above that use nucleons and nuclei in searches for violations of the Standard Model and searches for new physics, there are other more conventional measurements that have reached the level of precision required to probe potential BSM effects, provided that the theoretical Standard-Model predictions can become competitive with the experimental uncertainty. A prominent example is the single-$\beta$ decay of nucleons and nuclei, where measurements of rate at the 0.1\% level or better will provide competitive constraints for certain BSM scenarios compared with the LHC~\cite{Bhattacharya:2011qm,Cirigliano:2012ab,Gonzalez-Alonso:2018omy}. This will constrain the five dimensionless couplings parameterizing the various BSM couplings of the left-handed lepton current, namely a correction to the usual Standard Model left-handed quark current operator ($\epsilon_L$), a right-handed quark-current operator ($\epsilon_R$), and the three chirality-flipping scalar, pseudoscalar, and tensor quark operators ($\epsilon_{S,P,T}$)~\cite{Bhattacharya:2011qm,Cirigliano:2012ab,Gonzalez-Alonso:2018omy}. The matrix elements of the corresponding operators in the nucleon, namely isovector nucleon charges, provide key inputs into matching to the nucleon- and nuclear-level EFTs that are used to calculate the rate---quantities that lattice QCD can compute. 

As stated in a recent community whitepaper~\cite{Cirigliano:2017azj}, to compete with the LHC in constraining $\epsilon_{S,T}$ at the few $\times 10^{-4}$ level, which would probe effective scales of new physics close to 10 TeV, lattice-QCD calculations of scalar and tensor charges of the nucleon should reach $<10\%$ precision---a goal that has been already achieved in several studies~\cite{Aoki:2010xg,Green:2012ej,Bali:2014nma,Alexandrou:2017qyt,Gupta:2018qil}. Furthermore, a $0.2\%-0.3\%$ knowledge of the axial charge of the nucleon, $g_A$, will improve the model-independent bounds on possible right-handed currents~\cite{Bhattacharya:2011qm,Alioli:2017ces}. The lattice-QCD results on this quantity have started to reach few- to percent-level precision~\cite{Gupta:2018qil,Chang:2018uxx,Aoki:2021kgd}, but sub-percent-level precision will likely require substantial exascale resources in the coming years. Such precise calculations may also help resolve the current $\sim 4\%$ discrepancy in the results of neutron lifetime experiments using magnetic bottle and beam techniques~\cite{Berezhiani:2018eds}.

Lattice QCD can also help resolve the current puzzle in the apparent few-sigma deviation in the CKM unitary test arising from radiative corrections in $\beta$ decays~\cite{Seng:2018yzq}. For example, lattice-QCD constraints on the time-ordered product of weak and electromagnetic currents between neutron and proton at several momenta can calibrate the input into these calculations. Additionally, lattice QCD can directly compute the radiative corrections by computing the neutron to proton conversion in the presence of QED interactions. Both of these calculations are challenging but recent formal developments and numerical demonstrations in similar problems have paved the way to obtaining reliable estimates~\cite{Seng:2019plg,Giusti:2017dwk}.

\subsection{QCD calculations for dark matter}

\subsubsection{Dark-matter cross sections with nucleon and nuclei}
While gravitational signatures have confirmed the existence of dark matter in the universe, unraveling its nature remains an active area of research in HEP. In particular, directly detecting dark matter through its interactions with ordinary matter is the subject of great experimental efforts in the U.S. and elsewhere, as emphasized in a recent Snowmass whitepaper~\cite{Akerib:2022ort}. The lack of observation to date has greatly reduced the parameter space of  candidate models such as weakly interacting massive-particle (WIMP) models. It is expected that experiments planned over the next decade~\cite{Akerib:2022ort} will be able to probe the WIMP-nucleus cross section down to the ``neutrino floor", below which direct detection of dark matter will become much more challenging. Once again, accurate nuclear matrix elements are required to reliably convert limits on the cross sections to the WIMP mass or the parameters of other plausible models, and hence improve the confidence in the theoretical implications of the experimental outcomes. The various dark-matter candidates can be described systematically using Standard-Model EFT methods, and at dimension six and seven, these induce quark-bilinear operators of various Dirac structures. The role of lattice QCD is to constrain the matrix elements of these operators in nucleons and light nuclei that can then be incorporated into nucleon/nuclear EFTs that supplement nuclear many-body calculations in experimentally relevant targets such as xenon~\cite{Menendez:2012tm,Klos:2013rwa,Baudis:2013bba,Vietze:2014vsa,Hoferichter:2015ipa,Hoferichter:2016nvd,Hoferichter:2017olk,Fieguth:2018vob,Hoferichter:2018acd}.

Notable progress has been reported over the last decade in approaching few-percent-level precision in single-nucleon matrix elements of zero-momentum-transfer flavor-diagonal scalar, tensor, pseudoscalar, and axial currents with up, down, and strange quark contents as reported in the previous sections. Enhanced control of uncertainties in the determinations of various form factors will also be feasible in the coming years. As pointed out in a recent community whitepaper~\cite{Cirigliano:2017azj}, to go beyond a naive impulse approximation in which nuclear effects are ignored, a few-percent precision on the relevant matrix elements in light nuclei may be required using lattice QCD---a challenging goal that may only be achieved with sufficient exascale resources. Nonetheless, proof-of-principle calculations of the relevant matrix elements in light nuclei have started in recent years, aiming to shed light on the relative importance of nuclear effects with various currents, albeit at unphysically large quark masses~\cite{Chang:2017eiq}. 

There are many other possibilities for dark-matter interactions with Standard-Model particles beyond the scalar portal. For example, spin-dependent and velocity-dependent dark-matter interactions have been proposed, and constraining these models using experimental searches requires the knowledge a range of other nucleon- and nuclear-level matrix elements to be computed via lattice QCD, including PDFs, the progress in which is discussed in Sec.~\ref{sec:pdf}. Lattice-QCD determinations of the nonperturbative dynamics of dark-matter--nucleus scattering will, therefore, continue to contribute to the multi-scale matching between low-energy experiments and dark-matter models over the next decade.

\subsubsection{QCD properties and axion cosmology}

The QCD axion has been proposed as a dynamical solution to the strong CP problem, which in certain regions of parameter space can also serve as a dark-matter candidate;  see Ref.~\cite{Blinov:2022tfy} for further discussion. An alternative solution which was considered for some time was the possibility that the up quark is massless.  This hypothesis is now strongly disfavored by lattice-QCD calculation showing a nonzero up-quark mass~\cite{Aoki:2021kgd}, see also the more direct probe of topological mass contributions using lattice QCD in Ref.~\cite{Alexandrou:2020bkd}.  Because of its connection to the strong CP problem, certain properties of the axion (and therefore, potentially, of dark matter) are related to dynamical properties of QCD.

In particular, the axion mass $m_a$ can be related to the QCD ``topological susceptibility'' $\chi$, the second derivative of the free energy with respect to the CP-violating $\theta$ parameter.  This relationship is not sufficiently constraining to predict a unique value for $m_a$, but the dependence of $m_a$ on the temperature $T$ is relevant for predicting the relic density of axion dark matter.  Lattice calculations of the temperature dependence of $\chi(T)$ have already been carried out~\cite{Berkowitz:2015aua,Borsanyi:2015cka,Petreczky:2016vrs} at $\theta = 0$.  However, since the value of $\theta$ is dynamical in the early universe, understanding of axion cosmology in full generality requires knowledge of the dependence of $\chi$ on both $T$ and $\theta$.  Lattice simulation of QCD at finite $\theta$ suffers from a sign problem, and calculations of $\chi$ at arbitrary values of $\theta$ will require new approaches such as quantum simulation~\cite{Bauer:2022hpo}.

\section{Elucidating hadron structure and spectrum for High-Energy Physics
\label{sec:spec-struc}}
\subsection{Parton distribution functions
\label{sec:pdf}}
Hadrons are boosted to relativistic energies in collider experiments, such as in the LHC.  Hadron-hadron collisions produce a plethora of events in particle detectors that are analyzed to test Standard Model predictions and to search for signs of new physics beyond the Standard Model. At these energies, the scattering proceeds via the partonic constituents of the hadron, and hence a set of universal PDFs is required to predict the rates of the various processes. These distributions are best determined by ``global fits'' to all the available deep inelastic scattering and related hard scattering data~\cite{Martin:2009iq}. Unfortunately, these phenomenological extractions are subject to an inverse problem.  To solve this, one can resort to a factorization of the cross sections into hard scattering amplitudes, which can be calculated perturbatively, and nonperturbative PDFs. One then proceeds to introduce parameterizations of the PDFs that are fit to reproduce simultaneously the various cross sections. Since there is more than one such parametrization that can describe the data within uncertainties, this procedure introduces a systematic error. Additionally, there are significant uncertainties on the PDFs in the low-$x$ and large-$x$ kinematics, where $x$ refers to the fraction of the hadron momentum carried by the struck parton in the hadron's infinite-momentum frame. Another area where the global fits are unable to produce reliable predictions due to lack of pertinent experimental data is the flavor separation of unpolarized and polarized PDFs -- particularly in the strange sector, which plays an important role in precision electroweak physics such as the determination of the $W$-boson mass.\footnote{Furthermore, the precision with which the spin-dependent PDFs such as helicity and transversity PDFs can be inferred is limited relative to upolarized PDFs and these are needed to gain insight into the spin structure of the nucleon. Other important quantities such as transverse-momentum-dependent PDFs and generalized PDFs that allow a three-dimensional picture of the hadron to be elucidated have also remained largely unconstrained experimentally.} First-principles determinations of the PDFs based on QCD are, therefore, a critical task for the lattice-gauge-theory program as the LHC enters a high-precision era over the next decade~\cite{Lin:2017snn, Constantinou:2020hdm, Constantinou:2022yye}.

To understand the role of lattice QCD and its reach and limitations in the PDF program, it is important to first recognize that the lattice-QCD methodology in its present form does not allow for a direct evaluation of the PDFs, which are Fourier transformations of Minkowski-space light-cone correlations of quark and gluon fields in the hadron. This is because lattice-QCD calculations are preformed in Euclidean spacetime in order to enable a well-defined Monte Carlo sampling of the QCD path integral. In an operator product expansion of the hadron tensor, the DIS region can be described by matrix elements of local twist-2 operators at leading order which are accessible via the lattice-QCD method. These matrix elements give access to Mellin moments of distributions, from which the collinear PDFs can, in principle, be obtained from an inverse transform. Lattice QCD calculations of the lowest moment of PDFs have been successfully carried out over the years, see Refs.~\cite{Lin:2017snn, Constantinou:2020hdm} for reviews, but they are often limited to the first two moments due to a power-divergent operator-mixing issue resulting from reduced lattice symmetries compared with the continuum. New ideas to extend the reach of lattice QCD to higher moments are proposed~\cite{Detmold:2005gg,Davoudi:2012ya} and will continue to be tested in the coming years.

A full reconstruction of the PDFs requires many moments to enable a reliable inverse transform. Since only a few moments have so far been constrained, this method only gives access to momentum-space ``global'' information about partons, which makes it challenging to connect directly to a particular experiment in which particles of definite momentum are measured. This has promoted a range of alternative approaches that provide ``local'' information in momentum space by enabling direct access to the $x$-dependence of PDFs. A class of approaches that go beyond individual moments express the hadron tensor in the DIS region in terms of a Compton amplitude, and more generally current-current correlators in Euclidean space~\cite{Liu:1999ak,Liu:1993cv,Liang:2019frk,Aglietti:1998ur,Ji:2001wha,Chambers:2017dov,Sufian:2020vzb,Braun:2007wv,Ma:2017pxb,Radyushkin:2017cyf}. These methods typically use either a large momentum transfer to the currents that generate an operator product expansion or short-distance factorization directly in coordinate space. Lattice-QCD data on a range of coordinate-space correlations can be used to constrain PDFs through solving the inverse problem, either through a parametrization, neural network fit, or other Bayesian approaches such as the use of Backus-Gilbert and maximum-entropy methods~\cite{Karpie:2019eiq,Liang:2019frk,Cichy:2019ebf,DelDebbio:2020rgv,Zhang:2020gaj,Bringewatt:2020ixn,DelDebbio:2021whr}. Another increasingly popular approach which has generated promising results in recent years is the so-called quasi-PDFs method, which can be directly calculated on a Euclidean lattice~\cite{Ji:2013dva}.  This method does not involve solving an inverse problem, giving access to the PDFs via an ordinary time-independent momentum distribution function of a quark in the hadron.  The connection only holds in the infinite-momentum frame of the hadron, nonetheless an effective theory named large-momentum effective theory (LaMET) enables systematically characterizing the corrections to this picture in an in expansion in $\Lambda_{\rm QCD}^2/(xP)^2$ and $\Lambda_\text{QCD}^2/((1-x) P)^2$, where $P$ denotes the finite momentum of the hadron along the Wilson-line direction~\cite{Ji:2014gla,Ji:2020ect}. 

As argued in a Snowmass whitepaper~\cite{Constantinou:2022yye}, computing the matrix elements relevant for parton physics from lattice QCD faces a number of specific challenges: i) The need for a large momentum necessary in hadron/Compton-tensor currents or in hadron states in the quasi-PDF approach leads both to higher excited-state contamination and to an exponentially degrading signal-to-noise ratio in the correlation functions as a function of the time separation between the hadronic source and sink. Other scenarios with a poor signal are gluonic observables, correlation functions at large distances, and multi-variable matrix elements such as GPDs and TMDs. ii) Relevant to approaches that involve Wilson-line operators is precise renormalization, since these operators have linear divergences requiring high-precision control of UV physics. iii) The long-range correlations in coordinate space are necessary to reconstruct the PDFs from quasi-PDFs at a full range of $x$ values but due to confinement, such correlations decay exponentially with distance and require unprecedented precision to isolate them. iv) In extracting $x$-dependent PDFs in short-distance factorization approaches, reliably estimating systematics associated with the inverse problem is challenging given the discrete nature of data with finite statistics.

The lattice-QCD program in parton physics will be pushing forward in the upcoming years to control excited-state contamination, operator matching and renormalization, continuum extrapolation, and quark-mass and finite-volume extrapolations. Calibration with experimental data when present as well as benchmarking  moment-based and nonmoment-based methods against each other will also be pursued. It is encouraging that even at this early stage, some lattice-QCD calculations, particularly for spin-dependent quantities, are competitive with experimental data. This has led to a hybrid QCD global analysis combining lattice and experimental data~\cite{Lin:2017stx,Cichy:2019ebf,Bringewatt:2020ixn,DelDebbio:2020rgv,Hou:2022sdf}. According to the Snowmass whitepaper, if the agreement is met for both unpolarized and helicity isovector PDFs with 5\% precision, which is an achievable goal with exascale resources, lattice-QCD predictions can be used with high confidence for the transversity PDFs, as well as for PDFs in regions of $x$ or for parton flavors where experimental constraints are currently poor.

\subsection{QCD exotica and resonance physics
\label{sec:spectrum}}
The quest for establishing the existence of hadronic states that do not fit in the conventional organization of meson and baryon states continues to this date~\cite{Lebed:2016hpi,dzierba2000search}. Exciting experimental evidence for the tetraquark, $q\bar q q \bar q$, and pentaquark, $q\bar q q q q$, states has emerged in recent years at the LHCb, Belle, BESIII, and elsewhere~\cite{Belle:2011aa,BESIII:2013ris,LHCb:2021vvq,LHCb:2019kea}. In fact, approximately thirty exotic hadrons have been discovered and most of them contain at least one heavy quark~\cite{exotics}. The nature of these color-singlet states, e.g., whether they are compact quark bound states or molecular states of two hadrons (or perhaps a superposition of these or other possibilities), generally remains inconclusive~\cite{Brambilla:2019esw}. Additionally, the existence of hybrid hadrons, those that include valence gluons, is postulated, and such states are being searched for in various hadronic-physics facilities such as at Jefferson Lab~\cite{burkert2018jefferson,Dobbs:2017vjw}. Finally, both the neutrino-nucleus cross sections for neutrino-oscillation experiments and a number of hadronic decays measured at the LHC that hint new physics involve final-state resonances. As the QCD spectrum is generated by nonperturbative dynamics, first-principles calculations based in QCD using the lattice-QCD technique is the most rigorous path to theoretical predictions for such exotic states and resonances.

Presently, many of the states that are stable under strong interactions are determined with lattice QCD at a sub-percent statistical precision, with all systematic uncertainties quantified~\cite{Aoki:2021kgd}. Several calculations even incorporate small isospin-breaking effects due to QED and the difference in the light-quark masses~\cite{Aoki:2021kgd}. The associated energies are obtained by spectral decomposition of the two-point Euclidean correlation functions constructed from one or a set of operators that overlap with the desired states. The extracted energies can be reliably extrapolated to infinite volume since the mass of a stable hadron in a finite volume receives exponentially small corrections. However, the majority of states observed in experiment are not eigenstates of the QCD Hamiltonian, but rather resonances in the pertinent multi-hadron channels. This means that the determinations of a resonance's mass, decay width, and structure properties such as its form factors depend upon the calculation of a scattering amplitude in the relevant kinematic range. As there are no asymptotic states in a Euclidean finite lattice, the notion of a scattering S-matrix is ill-defined, hence indirect methods are sought to access S-matrices with lattice QCD.

A powerful methodology based on L\"uscher's work~\cite{luscher1991two,Luscher:1986pf} and generalizations gives access to scattering amplitudes of two-hadron elastic channels, of multiple coupled two-hadron inelastic channels, and of three-hadron channels including coupled two- and three-hadron processes, see Refs.~\cite{Briceno:2017max,Hansen:2019nir,Brambilla:2019esw,Mai:2022eur} for recent reviews. Development of these formalisms, along with extensions to local and nonlocal transition amplitudes (see Ref.~\cite{Davoudi:2020ngi} for a review) and corrections due to Coulomb interactions~\cite{Beane:2014qha,Christ:2021csx}, mark significant formal milestones in the lattice-QCD community in recent years. High-precision low-lying finite-volume spectra in a range of flavor channels are obtained using the lattice-QCD method and these are used as inputs into the mapping formalisms. The application of these formalisms has led to impressive results in recent years involving constraints on multiple two-hadron coupled channels in both light- and heavy-quark sectors, vector and scalar resonances, resonances involving three-hadron channels, and even resonance form factors and transition rates; see e.g., Refs.~\cite{Briceno:2017max,Brambilla:2019esw,Mai:2022eur} for recent reviews.

Despite this progress, the QCD resonance physics program still needs to overcome a number of challenges to be able to reach further above the strong-interaction thresholds, as identified in a Snowmass whitepaper~\cite{Bulava:2022ovd}. While L\"uscher's method for a single two-hadron elastic channel provides a straightforward one-to-one mapping between scattering amplitudes and finite-volume energies, for multi-channel or higher-body scattering this one-to-one mapping is lost, and the determination of the full scattering amplitudes often requires a parametrization of the amplitude or associated kernels. Abundant and precise energy eigenvalues in a given kinematic range are needed to robustly constrain these multi-parameter forms, with systematic uncertainties that need to be carefully quantified. As the calculations move toward physical values of the light quark masses, the multi-hadron thresholds move towards lower energies and the number of kinematically allowed hadronic channels that need to be included in the determination of scattering amplitudes is increased substantially, making it much more challenging to achieve the desired constraints. Over the next decade, the community will seek novel methods to circumvent some of these challenges. For example, it would be valuable to get access to higher-lying lattice-QCD spectra without the need to constrain the lower-lying energies in the process. Furthermore, novel techniques that do not rely on L\"uscher's approach can be explored, such as indirect evaluation of the amplitudes via an inverse transform from Euclidean to Minkoswki space under controlled conditions, as proposed in Ref.~\cite{Bulava:2019kbi}. 

In light of the challenges, what is the best way to proceed to complement the experimental and theoretical campaigns in the coming years? Highlighting the community consensus~\cite{Bulava:2022ovd}, several important directions can be enumerated:
\begin{itemize}
\item[--] First, it is necessary to identify channels and energy regions that feature certain exotic hadrons that can be reliably investigated with both lattice QCD and experiment. These provide clear targets both for validation of lattice-QCD methods, between lattice-QCD groups and against experimental information, and for reliable predictions that offer quantifiable systematic uncertainties.

\item[--] Second, it is important to note that the lattice-QCD studies of resonances can provide insights into the internal structure of the exotics in a way that is not accessible to experimental probes. For example, if a hadron mass is seen to remain near the threshold as the quark mass is varied, this could be interpreted as an indication of a sizable molecular component~\cite{RuizdeElvira:2017aet}. The structure of resonances can also be explored by evaluating the matrix elements of the currents that probe the charge or energy densities of a given flavor in momentum or position space, but the multi-particle nature of resonant states must be accounted for in such studies. 

\item[--] Third, lattice-QCD resonance physics has important applications in electroweak physics. With the current stage of developments and with sufficient computing resources, lattice QCD should be able to extract the electroweak transitions of nucleons to its resonant excitations, providing input to the resonance-production region of the neutrino-nucleus scattering cross section, as discussed in Sec.~\ref{sec:nu-nucleus}. Furthermore,  lattice-QCD studies of the weak decay $B \to K^*l^+l^-$ to an elastic resonance $K^* \to K\pi$ and any lepton pair $l = e,\mu,\tau$ will be valuable given that the experimental measurements hint to a possibility of new physics when comparing rates for various leptons in the final state~\cite{LHCb:2019hip,LHCb:2021trn}. Similarly, as the experimental rates for the decay $B \to D^*l \bar{\nu}$ to unstable $D^* \to D \pi$ indicate the violation of the lepton-flavor universality~\cite{BaBar:2013mob,Belle:2016dyj,LHCb:2015gmp}, lattice-QCD studies of this process should be given priority.
\end{itemize}
%

\subsection{Multi-nucleon systems and synergies between HEP and Nuclear Physics}
HEP experiments that are searching for new physics in rare processes often use large atomic nuclei to enhance signal rates, among other reasons as discussed in Sec.~\ref{sec:FS}. Furthermore, the physics output of long-baseline neutrino experiments depends on the knowledge of neutrino-nucleus cross sections to analyze and interpret the oscillation signals. This fact highlights the role of understanding and constraining Standard-Model and BSM interactions in nuclear media, and implies that only a coherent theory effort involving physicists in both high-energy and nuclear physics disciplines can lead to reliable results for the experimental program. To infer if an experimental measurement is consistent with the predictions of a proposed BSM scenario relies on performing a renormalization-group matching from high scales (TeV scale) down to the the low-energy nuclear-level scale (MeV scale). In particular, nuclear many-body theorists evaluate matrix elements of pertinent currents in experimentally relevant nuclei, nuclear EFTs construct and organize the \emph{ab initio} nucleon-level interactions and currents in the few-nucleon sector for use in larger nuclei, lattice-QCD physicists determine the quark- and gluon-level matrix elements in light nuclei to constrain the EFTs, and high-energy physicists match the matrix elements in the high-scale scenario down to the QCD scale, making use of factorization of perturbative and nonperturbative physics. An important role of the lattice-QCD program is, therefore, to go beyond constraining single-nucleon observables, and to provide critical input on the matrix elements in the two- and few-nucleon sectors.

Lattice-QCD calculations of baryonic systems are more challenging than their mesonic counterparts, due to the more complex quark-level construction of the correlation functions as well as exponential degradation of the signal resulting from lighter-than-baryon degrees of freedom that contribute to the noise correlator. Nonetheless, first lattice-QCD results for spectra of light nuclei and hypernuclei with atomic number $A<5$ have been reported~\cite{NPLQCD:2012mex,Yamazaki:2012hi,Yamazaki:2015asa,Inoue:2011ai}, albeit at unphysically large quark masses and without full control of systematic uncertainties such as discretization and potential excited-state effects. Under similar conditions, the first calculations of electromagnetic and weak reactions in the two- and three-nucleon systems have been performed, and the response of light nuclei to gluonic, scalar, vector, axial, and tensor external currents have been studied; see Ref.~\cite{Davoudi:2020ngi} for a review. The corresponding  low-energy coefficients of EFTs were constrained in several instances~\cite{Davoudi:2020ngi}. EFTs further used to calculate properties of larger nuclei, including Carbon and Oxygen, using many-body nuclear techniques~\cite{Barnea:2013uqa,Contessi:2017rww,Bansal:2017pwn}. These early calculations represent a demonstration of the matching program between QCD, EFT, and nuclear many-body theory, setting the stage for future complete lattice-QCD results at physical quark masses. 

The use of the variational technique in discerning the lowest-lying spectra of two-baryon systems has gained momentum in recent years with increased computational resources~\cite{Francis:2018qch,Horz:2020zvv,Amarasinghe:2021lqa}, and will further mature in the coming years. Additionally, the use of L\"uscher's methodology to constrain the scattering amplitude of two-baryon systems has generated valuable qualitative results at a range of quark masses as reviewed in, e.g., Refs.~\cite{Bulava:2022ovd,Detmold:2019ghl,Davoudi:2020ngi}, which have been used to constrain hadronic interactions in the relevant EFTs, see e.g., Refs.~\cite{Wagman:2017tmp,NPLQCD:2020lxg}. Nonetheless, these studies need to be refined when a full account of uncertainties is possible over the next decade. The first determinations of the three-hadron forces have become feasible using threshold expansions~\cite{Beane:2007es,Horz:2019rrn} or nonperturbative matching relations~\cite{Hansen:2020otl,Blanton:2021llb,Brett:2021wyd}, but it is likely that matching to EFTs in a finite volume will be the most straightforward approach as the atomic number increases or processes become more complex~\cite{Eliyahu:2019nkz,Detmold:2021oro,Sun:2022frr,Barnea:2013uqa,Contessi:2017rww,Bansal:2017pwn}.

In summary, given the evident role of accurate nuclear-level constraints based in QCD on the HEP program, the nuclear theory, lattice-QCD, and HEP communities will continue to push the frontier of this important program collectively over the next decade.

\section{
Reaching beyond QCD with lattice field theory
}

Many proposed theories of new physics beyond the Standard Model involve strongly-coupled gauge interactions in some way (in the present context, we focus primarily on strongly-coupled Yang-Mills theories which are asymptotically free.)  As with QCD in the Standard Model, strong coupling prevents the use of perturbation theory for predicting low-energy properties of the new sector.  Low-energy effective theories such as chiral perturbation theory or other phenomenological models can be used to gain some understanding, but here there is an additional challenge: without a concrete theory like QCD to match on to experimentally, these models appear to have a large number of free parameters and are difficult to validate.

Lattice gauge theory calculations provide a unique window on the physics of these proposed theories.  While extrapolation of QCD-based phenomenology may be difficult, in a lattice simulation we can directly alter the underlying characteristics of the theory: the number of colors $N_c$, the number of light fermion species $N_f$, or more exotic possibilities such as adding fermions in other gauge representations, charged scalars, or using gauge groups other than the special unitary group.  As a result of these additional variations, the parameter space for lattice studies of BSM physics is vast compared to lattice QCD.  Choosing a specific theory (number of colors, fermions, etc.) to study can often be motivated by matching on to a specific BSM model (the focus of Sec.~\ref{ssec:BSM} below.)  In other cases, we may attempt to use lattice calculations to study classes of theories and make general statements (Sec.~\ref{ssec:class}), for example studying the large-$N_c$ expansion.  Finally, lattice techniques can be used beyond the familiar territory of gauge theories, to study more exotic physical systems at strong coupling (Sec.~\ref{ssec:other_theory}).

Although the division of topics below is an effective way to organize this discussion, we emphasize that most lattice work in this overall area of research is exploratory and cross-cutting, and there are many overlaps and synergies.  A lattice calculation targeting a specific BSM model also gives new information about the behavior of strongly-coupled field theory in general, even if the model chosen is later ruled out.  Likewise, general information about the parameter space of strongly-coupled theories and their dynamics can inform new ideas for BSM models.

\subsection{Strongly-coupled extensions of the Standard Model \label{ssec:BSM}}

\subsubsection{Composite dark matter}
Most of the visible matter in the universe is composite, hence it is natural to take into account the possibility of a composite dark sector.  Theories with a strongly-coupled hidden sector can naturally give rise to composite dark-matter candidates.  Such theories are common in BSM scenarios, including composite Higgs, neutral naturalness, and grand unification---but composite dark matter is also interesting and well-motivated in its own right.  Compositeness can lead to distinctive dark matter properties, such as stability through accidental symmetry (similar to the proton) and direct interactions with light and other Standard-Model force carriers at high energies (analogous to the neutron).  For a more detailed review of composite dark matter and connections to lattice, see Ref.~\cite{Kribs:2016cew}; for more on the outlook for lattice calculations in particular, see the USQCD whitepaper~\cite{USQCD:2019hee}.

Lattice calculations can provide several quantities of direct interest for composite dark matter models.  Form factors of composite states are required inputs for the interaction rates of dark matter with ordinary matter; the electromagnetic and Higgs interactions are typically dominant, so information about scalar and vector form factors is generally the most useful.  There have been several lattice studies of such interactions for dark baryons or mesons in specific theories~\cite{LatticeStrongDynamicsLSD:2013elk,LatticeStrongDynamicsLSD:2014osp,Appelquist:2015zfa,Francis:2018xjd,Kulkarni:2022bvh}.  The spectrum of bound states is also an essential model input that is readily provided by lattice calculation.  A generic prediction of composite dark matter models with Standard-Model charges is the existence of charged bound states that can be searched for directly at colliders~\cite{Kribs:2018ilo,Kribs:2018oad,Butterworth:2021jto}, providing indirect bounds on the dark matter mass if their mass ratio is known.

There are several physical questions of great interest in the context of composite dark matter which will require new calculations and, in some cases, the development of new methods in order to study them.  One important question relates to dark matter self-interactions, as reviewed in Ref.~\cite{Tulin:2017ara}.  Self-interactions may help to explain certain astrophysical tensions between observed structure and predictions of cold, collisionless dark matter, but they are also constrained by observations at cluster scales.  Composite dark matter scattering can naturally show the velocity dependence required to explain the small-scale anomalies while satisfying other bounds.  Lattice calculations of hadron scattering interactions using L\"{u}scher's method~\cite{Luscher:1990ux} have already been carried out for mesons in theories other than QCD~\cite{Appelquist:2012sm,Drach:2020wux,LatticeStrongDynamicsLSD:2021gmp,Drach:2021uhl}.  Extensions of these studies to baryon scattering using methods and code developed for the study of nuclear physics in QCD should be straightforward.  The possibility of ``dark nucleosynthesis'', the formation of larger dark nuclei in the early Universe, can also be informed by binding energy results from the exact same lattice calculations and may be explored further by lattice studies of binding of light nuclei in BSM theories in the coming decade.

Turning to early-universe cosmology, the study of dark hadron interactions can also be important for calculation of dark matter relic abundance.  However, such a calculation includes annihilation processes (including two-to-many inelastic annihilation), which are extremely challenging to study on the lattice and will require new developments in formalism over the coming decade.  Annihilation is also a crucial input for ``indirect'' detection of dark matter through detection of gamma rays or other Standard-Model products.  Another important feature of composite dark matter sectors in the early universe is the finite-temperature ``dark confinement'' phase transition.  Although this transition is a crossover in QCD, it may be strongly first-order in other confining theories.  A first-order transition provides departure from thermal equilibrium and could be a crucial ingredient in explaining baryogenesis; finite-temperature lattice calculations can determine which strongly-coupled theories have a first-order transition.  If a first-order thermal transition does occur, it can source a primordial gravitational wave signal~\cite{Schwaller:2015tja,Huang:2020crf,Reichert:2021cvs} that can provide a novel signature of dark matter.  Lattice calculations of the finite-temperature properties of the theory e.g., Ref.~\cite{LatticeStrongDynamics:2020jwi} across the transition can allow for quantitative predictions of the gravitational wave spectrum, so that the reach of future experiments can be estimated (or conversely, if a discovery occurs, knowledge of how the spectrum is related to the parameters of the underlying theory can allow the dark sector to be characterized directly from such a signal.)

\subsubsection{Composite Higgs}
In the decade following the Higgs boson's discovery, there have been no further signs of new physics at the electroweak scale from the LHC so far.  Despite this experimental success of the Standard Model, theoretical questions such as the Higgs naturalness problem~\cite{Craig:2022uua} remain unanswered.  New-physics models which appear near the electroweak scale can resolve the electroweak hierarchy puzzle and other outstanding questions in HEP.  Composite Higgs theories, in which Higgs-sector naturalness is addressed by positing a new strongly-coupled theory from which the Higgs appears as a bound state, remain a viable and interesting possibility for BSM physics~\cite{Contino:2010rs,Bellazzini:2014yua,Panico:2015jxa,Cacciapaglia:2020kgq,Banerjee:2022xmu}.  However, stringent constraints from the LHC point to a scale separation between the Higgs boson and other new composite states, which indicate that the most viable models are those in which the Higgs is a pseudo-Nambu-Goldstone boson (PNGB).  This includes models where it is identified as a PNGB associated with chiral symmetry breaking (like the pion in QCD), as well as models with a pseudo-dilaton \footnote{This case has been strongly informed by lattice studies of a near-conformal phase in many-fermion gauge theories, with proposals of a dilaton EFT following in the wake of lattice calculations that revealed the emergence of a light scalar state.  This phase and the development of dilaton EFT is discussed further in Sec.~\ref{sssec:conformal}.}, a PNGB associated with scale-symmetry breaking.

If the Higgs is indeed composite, because it must be well-separated from the confinement energy scale, the first experimental evidence will likely be fully described by a low-energy effective theory whose structure is dictated by symmetries, and thus does not require a lattice calculation to formulate.  However, such an effective theory has many (in principle, infinite) parameters, the low-energy constants (LECs).  These are not free parameters; as with chiral perturbation theory in QCD, all of the LECs are determined by strong dynamics and a small number of underlying parameters.  Lattice calculations of LECs from first principles can thus severely constrain the available parameter space of a composite Higgs model, allowing for more concrete predictions for experiments.  It is also possible for lattice calculations in unfamiliar regimes to reveal unexpected dynamical surprises that lead to the construction of novel effective theories, such as the dilaton EFT (discussed further in Sec.~\ref{sssec:conformal}, along with the challenges for lattice studies of such systems.)

For a specific composite Higgs theory, if the ultraviolet (UV) completion (i.e. the underlying strongly-coupled gauge interaction) is specified, then lattice calculations can provide a number of useful inputs to constrain the phenomenology~\cite{USQCD:2019hee}.  In terms of LECs, details can vary by theory but relevant physics targets of interest include the $S$ and $T$ parameters, contributions to the Higgs potential, and to the top Yukawa coupling.  The mass spectrum of bound states is also of interest, allowing for prediction of the scale at which new resonances will appear in experiment.  Further work by phenomenologists on classification of UV-complete composite Higgs models, as in Refs.~\cite{Ferretti:2013kya,Ferretti:2014qta}, can guide future lattice calculations to the most promising theories; conversely, lattice results combined with experimental bounds may disfavor certain models, isolating those which are most interesting to study phenomenologically.

Although there is much progress that can still be made with current lattice methods, some of the more interesting questions related to composite Higgs models are presently inaccessible.  Four-fermion interactions are typically an essential ingredient of composite Higgs models, for example as a way to give rise to Standard-Model fermion masses.  If these interactions are weakly coupled, their effects may be included perturbatively.  But numerical study of the more general case is obstructed by the appearance of a sign problem when a generic four-fermion operator is included in the lattice action.  Innovations related to the sign problem relevant for other areas of study in lattice QCD may be adapted here over the next decade in order to explore the general effects of such interactions.

\subsubsection{Other examples of strongly-coupled new physics}
Strongly-coupled new physics sectors can manifest in a number of other ways, distinct from the composite dark matter and composite Higgs proposals.  One framework which has attracted growing interest in recent years is the idea of neutral naturalness~\cite{Batell:2022pzc}.  In neutral naturalness models, a broken discrete symmetry relates the Standard Model to a ``twin'' or ``mirror'' hidden sector in order to address the Higgs naturalness problem.  This generically leads to the existence of a hidden-sector SU(3) gauge interaction, similar to QCD---but not identical, due to the required breaking of the mirror symmetry.

There are many ways in which lattice calculations can contribute to the study of neutral naturalness models.  Typically, the Higgs vev in the mirror sector is higher, so that some or all of the mirror quark masses will be heavier than their Standard-Model counterparts.  Lattice QCD results at unphysically heavy quark masses can be highly relevant, and there are many such results from older QCD calculations where extrapolation to the physical point was required; see Ref.~\cite{DeGrand:2019vbx} for a review.

Other composite models for which lattice studies may be useful, but for which there has been little work to date, include theories of composite right-handed neutrinos~\cite{Arkani-Hamed:1998wff,vonGersdorff:2008is,Grossman:2010iq,Davoudiasl:2017zws,Chacko:2020zze} and composite axions~\cite{Kim:1984pt,Randall:1992ut,Lillard:2018fdt,Cox:2019rro,Ardu:2020qmo} (where compositeness can provide a solution to the ``axion quality problem".)  It may be interesting to consider lattice calculations targeting these types of theories, which could indicate novel directions to explore in theory space compared to the current set of lattice studies.  Finally, there may be cross-cutting aspects of the phenomenology of various types of composite theories.  For example, dark showers~\cite{Albouy:2022cin} are a distinctive phenomenological signature of hidden sectors with confinement, and lattice studies of the heavy-quark potential, string breaking, and PDFs at unphysically heavy quark masses may inform studies of hidden-sector hadronization independent of the specific model of interest.

\subsection{Understanding the theory space of strong dynamics}
\label{ssec:class}

\subsubsection{Large-$N_c$ expansion}
Large-$N$ limits are recognized in many areas of physics as a useful way to simplify the description of a theory or phenomenon.  For SU$(N_c)$ gauge theories at strong gauge coupling, taking $N_c \rightarrow \infty$ gives an alternative way to expand perturbatively using $1/N_c$ as a small parameter. This has been used to give qualitative insights into QCD, even though $N_c = 3$ is not quite large enough for the expansion to be obviously convergent.  In the broader context of understanding strongly-coupled gauge theory, large-$N_c$ formulas can be a guide to the larger parameter space, providing important input for BSM model-building.  Lattice calculations can be used to validate large-$N_c$ scaling formulas and quantify the size of deviations from scaling.  Lattice results also allow direct estimation of the coefficients appearing in large-$N_c$ scaling formulas, allowing for more accurate estimates in different theories\footnote{Using lattice results to determine large-$N_c$ expansion coefficients is especially useful for sub-leading corrections of order $1/N_c$ or $N_f / N_c$ (where $N_f$ is the number of light fermions, discussed below) - such corrections can be extremely difficult to calculate from first principles, and are highly important for some processes e.g.~$K \rightarrow \pi \pi$ decay~\cite{Cirigliano:2011ny,Hernandez:2020tbc}}.  For detailed reviews of large-$N_c$ expansion in the context of lattice studies, see Refs.~\cite{Lucini:2012gg,Hernandez:2020tbc}.

In the well known 't Hooft limit~\cite{tHooft:1973alw}, $N_c \rightarrow \infty$ while the number of light fermions $N_f$ is held fixed.  The effects of the fermions generally decouple in this limit, and lattice studies of the corresponding pure-gauge theories have been carried out in substantial detail.  Detailed results for the finite-temperature phase transition and the spectrum of glueball masses at SU$(N_c)$ are available at multiple values of $N_c$, and good agreement with the predicted $N_c$ scaling is seen~\cite{Morningstar:1997ff,Lucini:2010nv}.  These existing lattice results have already been used as valuable input for a broad range of BSM models in which hidden-sector glueball states appear.  For such model applications, in addition to the glueball spectrum, it would be very interesting to explore glueball decay matrix elements and glueball scattering states in large-$N_c$ pure-gauge theory; although these are very challenging calculations with few results available currently, a systematic study will likely be feasible with advances in computing and methodology over the next few years.

Another interesting frontier in the study of large-$N_c$ gauge theories is to move away from the 't Hooft limit to the Veneziano limit, in which the ratio $N_f / N_c$ is held fixed as $N_c \rightarrow \infty$ so that the fermions do not simply decouple.  In analogy to early lattice QCD calculations, this can be compared to moving from the quenched limit (in which quark contributions are neglected) to the more challenging case of fully dynamical quarks.  Although theoretical calculations including fermion contributions are much more challenging, lattice calculations with dynamical fermions are relatively straightforward, and $N_f / N_c$ corrections to 't Hooft scaling formulas can be obtained phenomenologically by carrying out fits to lattice data.  Further study of $N_f / N_c$ corrections in large $N_c$ on the lattice~\cite{DeGrand:2016pur,DeGrand:2021zjw} may not only aid in making more reliable predictions for BSM models, but can also enable quantitative applications of large-$N_c$ results to QCD~\cite{Donini:2020qfu,Baeza-Ballesteros:2022azb}.

\subsubsection{Emergent conformal symmetry}\label{sssec:conformal}
Conformal field theories are a highly important and interesting class of theories in particle physics, as well as in other areas such as condensed matter physics.  These theories obey conformal symmetry, which implies the existence of scale invariance (i.e.~the theory is self-similar at all length/energy scales.)  Conformal theories are interesting in their own right, with deep connections to the physics of renormalization and quantum phase transitions, as well as to quantum gravity as the latter half of the AdS/CFT correspondence.  They are also of interest as key parts of certain BSM models, such as in composite Higgs models where approximate scale invariance allows for construction of realistic models which give rise to the Standard-Model fermion masses.

It is well known that within the space of Yang-Mills gauge theories coupled to fermions, there is an emergent infrared-conformal phase when the number of fermions is large enough; see Refs.~\cite{DeGrand:2015zxa,Svetitsky:2017xqk,Witzel:2019jbe,Drach:2020qpj} for recent reviews in the context of lattice simulations.  Although the existence of this phase was established with perturbation theory shortly after the discovery of asymptotic freedom in QCD~\cite{Caswell:1974gg,Banks:1981nn}, lattice calculations are required to explore the transition itself, which occurs at strong gauge coupling.  Despite significant efforts on lattice simulations of many-fermion theories\footnote{The majority of these efforts have been concentrated in SU$(3)$ with $N_f$ fermions in the fundamental representation, due to the ease of repurposing QCD software---but there are now many exploratory studies of infrared-conformal gauge theories where the number of colors and gauge representation are varied as well.},  the precise location and order of the conformal transition remains unknown to date. The study of these theories is extremely challenging; lattice simulations require several sources of explicit scale-symmetry breaking which can be difficult to control, and the presence of lattice-phase transitions at strong bare coupling has obstructed access to the infrared limit in theories near the transition.

Nonetheless, there have been significant discoveries from lattice studies of the conformal transition.  Spectrum calculations by multiple groups~\cite{LatKMI:2013bhp,LatKMI:2013usl,LatKMI:2014xoh,LatKMI:2016xxi,Athenodorou:2014eua,Fodor:2015vwa,Fodor:2016pls,Brower:2015owo,DelDebbio:2015byq,Hasenfratz:2016gut,Appelquist:2016viq,LatticeStrongDynamics:2018hun} have revealed the presence of a relatively light scalar bound state with $J^{PC} = 0^{++}$ quantum numbers in multiple theories, making it a candidate for a pseudo-dilaton associated with scale symmetry breaking.  The lattice results led to substantial theoretical work on formulation of a ``dilaton EFT'', which is able to explain the emergence of such a state near the conformal transition as scale symmetry is approximately restored.  Further lattice studies of the spectrum, pion scattering, and other observables in candidate theories such as SU$(3)$ with $N_f = 8$ can further inform theoretical studies and help to distinguish between different formulations of the dilaton EFT~\cite{Hansen:2016fri,Golterman:2016lsd,Appelquist:2017wcg,Appelquist:2017vyy,Appelquist:2019lgk,LSD:2018inr,Golterman:2020tdq,Golterman:2020utm}.

In the next decade, continued study of the conformal transition and the emergent CFTs that result must be a high priority for lattice field theory.  Beyond the study of the transition itself, the properties of the CFT such as anomalous dimensions of operators are of great interest for classifying and understanding the CFTs themselves, as well as for application to BSM models such as composite Higgs (where large anomalous dimensions for the fermion mass operator are an essential ingredient in generation of realistic quark masses.)  New methods for studying renormalization-group (RG) transformations, such as gradient flow RG~\cite{Fodor:2017die,Carosso:2018bmz,Hasenfratz:2019hpg}, can be helpful in enabling such calculations.  Moreover, the study of new RG techniques can feed back into lattice QCD, where they might be used for other applications~\cite{Hasenfratz:2022wll}.  Further advances in lattice actions to enable simulation at stronger coupling~\cite{Hasenfratz:2021zsl}, or radial quantization methods which exploit a conformal mapping of the theory onto a curved space~\cite{Brower:2012mn,Brower:2012vg,Brower:2014daa,Brower:2020jqj}, may enable much more accurate studies of the infrared-conformal limits and thus allow more precise answers about the conformal transition.

\subsection{Pushing the boundaries of particle theory}
\label{ssec:other_theory}
Going beyond simple extensions from QCD, there are other, more exotic theories that can also be studied nonperturbatively using lattice methods.  In addition to the examples listed here, it is worthwhile to consider how lattice results can contribute to more general studies.  For example, computing directly at strong coupling in specific theories may be helpful in understanding the nature of dualities and generalized symmetries~\cite{Cordova:2022ruw}.

Supersymmetry is of broad interest to the theoretical particle physics community.  Regardless of whether it is realized in Nature near the electroweak scale, supersymmetry is a key part of holography through the AdS/CFT correspondence, which has given essential insights into our theoretical understanding of quantum field theory and gravity.  

One supersymmetric theory of particular interest is $\mathcal{N} = 4$ super-Yang-Mills (SYM) theory; see Ref.~\cite{Catterall:2022qzs} for a whitepaper on lattice calculations in this theory and Ref.~\cite{Catterall:2009it} for further details.  Substantial work in recent years has gone into a lattice formulation of $\mathcal{N}= 4$ SYM, which can be done through a novel lattice formulation which preserves a subset of the continuum supersymmetry, allowing recovery of the full continuum limit with little to no fine-tuning.  Lattice calculations in this theory so far agree well with holographic predictions---surprisingly, even for small numbers of colors, even though the holographic correspondence holds in the large-$N_c$ limit.  Future work may include tests of S-duality, calculation of operators which can be tested against the conformal bootstrap, and numerical exploration of quantities corresponding to string-loop corrections via holography.

A possible feature of the emergent infrared-conformal phase in gauge-fermion theories is the presence of nontrivial ultraviolet fixed points; these would be examples of ``asymptotically safe'' systems \cite{Weinberg:1980gg,Bond:2016dvk,Litim:2014uca,Bond:2017suy,Bond:2017tbw}, which could be relevant to models of quantum gravity \cite{Litim:2006dx,Niedermaier:2006wt,Percacci:2007sz}.  If these limits of many-fermion theories do exist, they may be discovered and explored in the next decade by lattice studies.  In the same context of the infrared-conformal phase transition, there have also been potential hints from lattice results of a symmetric mass generation phase~\cite{Butt:2018nkn, Butt:2021koj, Hasenfratz:2022qan}.  The phenomenon of symmetric mass generation is gathering new interest in the context of both HEP~\cite{Tong:2021phe, Catterall:2022jky} and condensed matter physics~\cite{Wang:2022ucy}, and a concrete example of such a system which could be studied nonperturbatively could be very fruitful in understanding this mechanism.

\section{Advancing theory and computation}

\subsection{Advancing computational algorithms and software}
As experimental measurements become more precise over the next decade, lattice QCD will play an increasing role in providing the needed matching theoretical precision. Achieving the needed precision requires simulations with lattices with substantially increased resolution.  With finer lattice spacing comes an array of new challenges. They include algorithmic and software-engineering challenges, challenges in computer technology and design, and challenges in maintaining the necessary human resources.  

As a computational problem, lattice gauge theory is performed on structured Cartesian grids with a high degree of regularity and natural data parallelism. The approach formulates the Feynman path integral for QCD as a statistical mechanical sampling of the related Euclidean-spacetime path integral. 
The computing landscape at this time displays an exciting 
proliferation of competing computer architectures, and
the massive parallelism of lattice gauge theory is, 
in principle, amenable to Graphical Processing Units (GPUs) and possibly other acceleration.
This imposes a significant additional programmer overhead. The
most commonly-used packages receive sufficient 
investment to use the complete
range of modern accelerated supercomputers, and many of the largest
projects use allocations on Department of Energy's (DOE's) supercomputer resources.
This investment must continue to realize the scientific
goals of the community.
Interconnects are becoming an increasing bottleneck since accelerated computing nodes are becoming rapidly
more powerful while interconnect performance gains have not always matched pace. 

The Lattice QCD workflow is divided into two phases. First, 
a Markov Chain Monte Carlo (MCMC) sampling phase generates
an ensemble of the most likely gluon field configurations distributed
according to the QCD action.
The ensemble generation is serially dependent and represents a strong scaling
computational problem. 
On the largest scales, this becomes a halo-exchange
communication problem with a very large interconnect bandwidth requirement 
since the local data bandwidths vastly exceed those of inter-node communication.
In the second phase, hadronic observables are calculated
on each sampled configuration where many thousands of quark propagators (requiring inverting high-dimensional 'Dirac' matrices)
are calculated and assembled into hadronic correlation functions.
This has a high degree of trivial parallelism. 

Present algorithms for both path integral sampling and Dirac solvers display growing limitations as substantially greater ranges of energy scales are included in the problem, an algorithmic challenge called critical slowing down.
The development of numerical algorithms is a significant intellectual activity that spans physics, mathematics, and computer science.
The U.S. HEP program is presently focused on the
domain-wall and staggered approaches,
and multigrid algorithmic  
gains for staggered and domain-wall-fermion discretizations
are a critical open research activity. 
A second algorithmic direction is the critical slowing down of
MCMC algorithms. 
These are being studied under Exascale Computing and
SciDAC projects and such support is critical to 
progress in the field.
The simulations at finer lattice spacing and larger volumes required to realize the community goals introduce new challenges~\cite{Boyle:2022ncb}:
\begin{itemize}
    \item[--]{``Meeting them requires new algorithmic research, novel computer hardware design beyond the exascale, improved software engineering, and attention to maintaining human resources.''}
\end{itemize}
A recent Snowmass whitepaper~\cite{Boyle:2022ncb} estimates that:
\begin{itemize}
\item[--]{``The simulation goals [of flavor physics, nucleon structure, neutrino scattering, and BSM physics] clearly demonstrate a need
for computers at least 10x more capable than the coming Exaflop
computers during the Snowmass period. Since the performance is required to be delivered on a real-code performance basis...
more than an order of magnitude improvement, perhaps, from both
algorithms and computing are required.''}
\end{itemize}

Commonly used lattice-QCD software has been supported
by the DOE SciDAC and Exascale Computing Projects.
These include \textsc{Grid}~\cite{Boyle:2022nef,Boyle:2016lbp,Boyle:2017gzg}, \textsc{MILC}~\cite{DeTar:2018pyj,Gottlieb:2000tn}, \textsc{CPS}~\cite{Jung:2014ata}, \textsc{Chroma}~\cite{Edwards:2004sx}, and \textsc{QUDA}~\cite{Clark:2009wm}.
This has enabled major packages to support the most advanced GPU accelerated HPC computers using software interfaces such as
HIP, SYCL, and CUDA APIs in addition to giving good performance
on several CPU SIMD architectures. 
Newer interfaces like OpenMP 5.0
offload and C++17 parallel STL are planned to
be adopted as and when appropriate.

For smaller projects, especially where rapid development and programmer
productivity are at a premium, it is better and more cost effective
in terms of human effort to maintain access to a range of CPU resources.
USQCD institutional clusters at Brookhaven National Laboratory, Fermilab, and Jefferson Laboratory have
been instrumental in supporting the significant number of smaller and exploratory projects that would not
achieve the return on investment to justify bespoke software development for multiple architectures.

Reference~\cite{Boyle:2022ncb} notes the heavy reliance on HPC places a particularly large dependence on highly-tuned 
bespoke software in this field.
It was noted as important that flexible high-performance software continues to be developed for a diverse range of architectures that tracks the DOE computing program.

\subsection{Machine-learning applications in lattice field theory }
Machine learning (ML) methods have seen an explosion of interest in recent years throughout the computational community, and broadly in theoretical particle physics~\cite{Gupta:2022vhe}.  In lattice field theory, there has been rapidly growing exploratory work in applying ML to all aspects of the numerical process: configuration generation, measurement of observables, and physics analysis~\cite{Boyda:2022nmh}.  Although ML methods have yet to be widely adopted for large-scale lattice QCD calculations, they hold great potential for algorithmic breakthroughs, ranging from significant speed-up of configuration generation to solutions to perform currently intractable calculations which suffer from sign problems or ill-posed inverse problems.

Important similarities between ML and lattice-field-theory methods and practices give a unique opportunity for collaborative development of methods at the intersection of the two fields.  The mathematical frameworks for algorithms in both ML and lattice field theory are based heavily on statistics and linear algebra, allowing researchers to readily transfer specialized knowledge in one field to similar problems in the other.  There are also similarities in computational architectures well-suited to both fields, highly parallel calculations of numerical linear algebra. ML applications tend to require lower precision, but with the use of mixed-precision methods lattice calculations can also exploit low-precision calculations (e.g. on graphical processing units) to accelerate their calculations.  The ability of common hardware to support cutting-edge calculations in both ML and lattice field theory makes rapid progress at the intersection of the two much easier to attain.

Beyond application of existing ML methods to lattice field theory, the development of new approaches to ML inspired by physics problems may feed back into innovations relevant more broadly for ML uses in academia and in industry.  A prime example is the incorporation of symmetries into ML architectures~\cite{Bogatskiy:2022hub}.  Although the use of symmetry is not unique to physics problems (most famously, the use of convolutional neural networks which have translational symmetry has been enormously successful in image processing), symmetry has always been a guiding principle in theoretical physics and the symmetry properties of physical systems are well studied.  Incorporation of symmetry principles into the architecture of a ML network can greatly improve performance, as demonstrated for example in the use of SU$(N)$ gauge equivariance in ML-based sampling algorithms~\cite{Kanwar:2020xzo,Boyda:2020hsi}.

There are significant obstacles remaining for widespread and routine deployment of ML methods in lattice-field-theory calculations.  Incorporation of fermions into methods for flow-based configuration generation has been demonstrated in toy problems~\cite{Albergo:2021bna,Albergo:2022qfi}, but new methods will be needed for estimation of the fermion determinant in order to apply the method to realistic lattice-QCD simulations.  In general, interpretation of results and statistical control of uncertainty estimates is a concern for any use of ML, due to the ``black box'' nature of many ML methods. There are methods available to ensure that ML-obtained lattice-field-theory results are statistically rigorous, but this is an important constraint on new techniques.  Finally, successful development of methods at the intersection of ML and lattice field theory requires both scientists and software engineers with expertise in both domains. Therefore, workforce development and support will be key, particularly in a field in which there is substantial demand for early-career researchers outside of academia.

\subsection{Hamiltonian-simulation methods and quantum computation}
The lattice-field-theory technique conventionally relies on Monte Carlo importance sampling methods to evaluate QCD path integrals. Such a sampling is made possible by a Wick rotation to Euclidean spacetime. However, this statistical feature has led to limited progress in several problems including finite-density systems and real-time phenomena such as scattering processes, except for those at low energies and low inelasticities that can be addressed by indirect methods, see Sec.~\ref{sec:spectrum}. A Hamiltonian-simulation approach does not encounter such issues, but the size of the required Hilbert space scales exponentially with the system size, rendering the simulations impractical on even largest supercomputers. In recent years, Hamiltonian-simulation methods based on tensor networks have significantly advanced~\cite{orus2014practical,orus2019tensor,Cirac:2020obd}, targeting generally low-dimensional theories and systems without volume-law entanglement. The progress in the applications of tensor networks to lattice gauge theories in both the Hamiltonian and path-integral formulations will continue over the next decade, as discussed in a Snowmass whitepaper~\cite{Meurice:2022xbk} and recent reviews~\cite{Banuls:2019rao,Meurice:2020pxc}. Nonetheless, more general Hamiltonian-simulation methods are needed, particularly pertinent to QCD and for real-time situations that exhibit an entanglement growth.

A natural method for realizing Hamiltonian simulation of quantum field theories is quantum simulation. Quantum simulation refers to simulating a complex quantum system using a more controlled quantum system, often as table-top experiments. The dynamics of an analog quantum simulator can be engineered to closely follow those of the simulated theory, and can exhibit a variety of degrees of freedom, from two- and higher-dimensional spins, to fermions, bosons, and even continuous variables. On the other hand, a digital quantum simulator is agnostic to the underlying physical architecture, and implements a set of universal and elementary operations on an array of two-(few-)state quantum units, i.e., qubits (qudits). The two modes of the simulator can also be combined to achieve more adaptability and efficiency. In the context of simulating quantum field theories of Nature and beyond, as is argued in a Snowmass whitepaper~\cite{Bauer:2022hpo}, continuous research and innovation in theory, algorithms, hardware implementation, and co-design are critical to advance this field to the level that is needed for addressing the physics drives of the HEP program.

In particular, upon development of theory and algorithms, and provided robust large-scale quantum hardware becomes a reality in the coming years,  quantum simulation has the potential to enable~\cite{Bauer:2022hpo}: i) computations of full scattering processes for high-energy colliders, as well as first-principles calculations of PDFs and other QCD matrix elements, ii) simulations of neutrino-nucleus scattering cross sections crucial for supplementing the experimental programs such as DUNE and other neutrino processes for astrophysics, iii) studies of nonequilibrium dynamics in particle collisions, in inflationary phase of the universe, in CP-violating scenarios, and for models of dark matter, and iv) elucidating bulk gravitational phenomena and accessing quantum gravity in table-top experiments.

All these problems, in one form or another, involve simulating quantum (fundamental or effective) field theories, hence lattice gauge theorists, given their computational expertise, are in a prime position to leverage the potential benefits of the growing quantum-simulation hardware and its algorithm/software/compiler/user ecosystem. In the next decade, several interconnected problems will be addressed to make HEP problems accessible to quantum simulators~\cite{Bauer:2022hpo}:
\begin{itemize}
\item[--] Efficient formulations of gauge-field theories are needed within the Hamiltonian framework. A main requirement is to turn the infinite-dimensional Hilbert space of quantum field theories into a finite-dimensional one, via discretizing the space in a lattice formulation (as a common but not unique option), and digitizing the field values. More research is required to determine the pros and cons of each formulation developed, particularly for gauge theories of the Standard Model. Moreover, finding optimal ways to protect or utilize the symmetries of the underlying theories, especially gauge symmetries, is an active area of research. Finally, systematic uncertainties of quantum simulations of quantum field theories, including those associated with the truncations imposed on the infinite-dimensional bosonic-field Hilbert spaces, and issues related to time digitization (in digital implementations), finite-volume effects, and continuum limit and renormalization, need to be investigated carefully. 
\item[--] Algorithmic research for both digital and analog simulations is required, rooted in theoretical understanding, and taking advantage of physics inputs such as locality, symmetries, and gauge and scale invariance when present. For digital schemes, low-overhead and efficient encodings of degrees of freedom to qubits, and algorithms with tight and rigorous error bounds are needed to come up with realistic resource estimates for the desired simulations. Furthermore, concrete protocols are needed for preparing nontrivial initial states and for measuring the outcome relevant for a range of observables, from scattering amplitudes and structure functions, to identifying phases of matter and its evolution under extreme conditions, through quantum-information measures and other methods.
\item[--] Since certain problems might be more natural to implement on analog quantum simulators, algorithms need to be developed to utilize analog quantum simulators for HEP problems. This amounts to understanding how one can map given quantum field theories onto a variety of analog quantum simulators, including atomic, molecular, optical, and solid-state systems, each with distinct intrinsic degrees of freedom, native interactions, and connectivity properties. Enhanced modalities and advanced quantum-control capabilities must be co-developed between lattice gauge theorists and hardware developers to enable simulating QCD and other complex field theories in the coming years.
\end{itemize}

To ensure the feasibility of the simulation approaches, prevent a wide gap between theorists’ proposals and experimental realities, and tighten the theoretical algorithmic scalings by supplementing empirical observations, implementation and benchmarking using the near-term noisy intermediate-scale quantum hardware will continue to be a critical path forward for the HEP community. Collaboration among universities, national laboratories, and private companies is essential in this growing multidisciplinary field. Lattice gauge theorists will likely engage in a careful study in the coming years to determine whether industry-developed quantum hardware satisfies the needs of the HEP community or special-purpose hardware may need to be co-designed, following the development and application of special-purpose HPC hardware/software in the lattice-QCD research in the past~\cite{boyle2005overview,boyle2004qcdoc}; see also the Snowmass whitepaper on quantum computing systems and software for HEP~\cite{Humble:2022vtm}. Additionally, lattice gauge theorists will find ways to leverage the current advancements on classical computation of field theories and augment them with quantum-computing routines. Quantum computing and quantum technologies constitute another area with high demand for experts outside academia, providing alternative career paths for young trainees in this field, but necessitating strategies for growing and retaining a diverse and engaged workforce at the intersections of QIS and lattice field theory.

\phantomsection
\addcontentsline{toc}{section}{Acknowledgment}
\section*{Acknowledgments}
This work was supported by the U.S.\ Department of Energy, Office of Science, Office of High-Energy Physics under grants DE-SC0021143 (ZD), DE-SC0010005 (ETN), DE-SC0010339 (TBl), DE-SC0015655 (AE, ATL), DE-SC0011941 (NC), DE-SC0009913 (S.~Meinel), DE-SC0019139 and DE-SC0010113 (YM), DE-AC02-05CH11231 (CB), DE-SC0010339 and DE-SC0021147 (LCJ); by the U.S.\ Department of Energy, Office of Science, Office of Nuclear Physics under grants DE-SC0011090 (WD, DCH, WIJ, PES), DE-SC0021006 (WIJ, PES), and DE-SC0012704 (S.~Mukherjee); and by the National Science Foundation under QLCI grant OMA-2120757 (ZD), EAGER grant 2035015 (PES), grants PHY 1653405 and PHY 2209424 (HL), and by grant PHY20-13064 (CD).  

ZD is also supported by the U.S.\ Department of Energy, Office of science, Office of Advanced Computing Research, Quantum Computing Application Teams program under fieldwork proposal number ERKJ347.  WD is also supported by the SciDAC4 award DE-SC0018121.  PES is also supported by the National Science Foundation under Cooperative Agreement PHY- 2019786 (The NSF AI Institute for Artificial Intelligence and Fundamental Interactions, http: //iaifi.org/).  TBh is partly supported by the U.S. Department of Energy, Office of Science, Office of High Energy Physics under Triad National Security, LLC Contract Grant No. 89233218CNA000001 to Los Alamos National Laboratory.  ML was supported in part by the Exascale Computing Project (17-SC-20-SC), a collaborative effort of the U.S. Department of Energy Office of Science and the National Nuclear Security Administration, in particular its subproject on Lattice QCD software and algorithm development.  SP is supported by ARRS project  P1-0035.  HL is also supported by the  Research  Corporation  for  Science  Advancement through the Cottrell Scholar Award.  CB is also supported by the U.S. Department of Energy's (DOE's) Office of Science under the Quantum Information Science Enabled Discovery (QuantISED) for High Energy Physics grant number KA2401032.  This document was prepared using the resources of the Fermi National Accelerator Laboratory (Fermilab), a U.S.\ Department of
Energy, Office of Science, HEP User Facility.
Fermilab is managed by Fermi Research Alliance, LLC (FRA), acting under Contract No.\ DE-AC02-07CH11359.

\bibliographystyle{unsrt}
\bibliography{main.bib}

\end{document}